\documentstyle[aps,eqsecnum,epsf]{revtex}

\begin{document}
\thispagestyle{empty}

\begin{flushright}
JLAB-THY-99-12 \\
\today  
\end{flushright}

\begin{center}
{\Large \bf  Evolution and Models for Skewed  Parton Distributions }
\end{center}
\begin{center}
{I.V. MUSATOV$^a$ and A.V. RADYUSHKIN$^{a,b}$ \footnotemark
}  \\
{\em $^a$Physics Department, Old Dominion University, Norfolk, VA 23529, USA}
 \\ 
{\em $^b$Jefferson Lab, Newport News, VA 23606, USA}
\end{center}
\vspace{2cm}

\footnotetext{Also Laboratory of Theoretical  
Physics, JINR, Dubna, Russian Federation}
\date{\today}

\begin{abstract}

We discuss the structure of the 
``forward visible''  (FW) parts of 
double and skewed distributions 
 related to usual distributions
through reduction relations.  
We use  factorized models for
double distributions (DDs) $\tilde f(x,\alpha)$ in which one 
factor coincides with the usual (forward)
parton distribution  and another 
specifies the profile characterizing the
spread  of the longitudinal momentum transfer. 
The model DDs   are used to construct 
skewed parton distributions (SPDs). 
For small skewedness, the FW parts of 
SPDs $ H(\tilde x , \xi)$ can be obtained 
by averaging  forward  parton
densities  $f(\tilde x-\xi \alpha ) $  with the weight
$\rho (\alpha)$ coinciding with the profile function
of the double distribution $\tilde f(x,\alpha)$ 
at small $x$. 
We show that  if the 
$x^n$ moments $\tilde f_n (\alpha)$ of 
DDs have the asymptotic $(1-\alpha^2)^{n+1}$
profile, then the $\alpha$-profile of $\tilde f (x,\alpha)$ 
for small $x$ is completely determined  
by   small-$x$ behavior of the usual parton distribution.
We demonstrate that, for  small  $\xi$, the 
model with asymptotic profiles for $\tilde f_n (\alpha)$
is equivalent 
to that proposed recently by Shuvaev et al.,
 in which  the Gegenbauer moments of SPDs do not depend on $\xi$.
We perform a numerical investigation 
of the evolution patterns of SPDs and gave interpretation
of the results of these studies within the 
formalism of double distributions.

\end{abstract}

\section{Introduction}

Applications of perturbative QCD to
deeply virtual Compton scattering (DVCS) 
and hard exclusive electroproduction 
processes \cite{ji,compton,npd,cfs,drm} 
involve nonforward matrix
elements $\langle p-r \,  |     \,   
{\cal O}(0,z)  \,  |     \, p \rangle \,  |     \, _{z^2=0}$ 
of quark and gluon light-cone  operators. 
They     can be parametrized by 
two basic  types of nonperturbative functions. 
The   double distributions (DDs) $F(x,y;t)$ \cite{compton,npd,ddee,sssdd}   
specify the Sudakov
light-cone ``plus'' fractions $xp^+$ and $yr^+$ 
of the initial hadron momentum $p$ and the momentum transfer $r$ 
carried by the initial parton.   
Treating the proportionality  coefficient 
$\zeta$ as an 
independent parameter  one can   introduce 
an alternative description in terms
of the   nonforward parton distributions (NFPDs) 
${\cal F}_{\zeta}(X;t)$ 
with $X=x+y \zeta$ being the total 
fraction of the initial hadron momentum 
taken  by the initial  parton.
The shape of  NFPDs  explicitly
depends on the parameter $\zeta$ characterizing the {\it skewedness}
of the relevant nonforward matrix element.
This parametrization of  nonforward matrix 
elements  by ${\cal F}_{\zeta}(X;t)$
is similar to that proposed  by 
X. Ji \cite{ji} who introduced 
 off-forward parton distributions (OFPDs) $H(\tilde x,\xi;t)$
in which the parton momenta and  the skewedness
parameter $\xi\equiv r^+ / 2 P^+$ 
are measured in units of the average 
hadron momentum $P=(p+p')/2$. 
  OFPDs and NFPDs \cite{npd,cfs}
can be   treated 
as particular forms  of {\it skewed } parton 
distributions (SPDs).
One can also introduce the version of  DDs (``$\alpha$-DDs'' \cite{sssdd}) 
in which the active parton momentum is  written in terms of symmetric 
variables  
$k= xP + (1+\alpha) r/2$.    

In our approach, DDs are  primary
objects producing SPDs after an appropriate integration.
In refs. \cite{ddee,sssdd} it was   shown that  
using the    the
support and symmetry properties of DDs,  
one can easily establish important features of SPDs
such as nonanalyticity at border points $X=\zeta,0$  
[or $\tilde x = \pm \xi$], polynomiality of their $X^N$ and 
$\tilde x^N$ moments in skewedness parameters $\zeta$ and $\xi$, etc.
 
The physical interpretation
of DDs $F(x,y;t=0)$ [or $f(x,\alpha;t=0)$ ]
 and their relation to
the usual parton densities $f(x)$  
 suggests   that  the $x$-profile of DDs $F(x,y), f(x,\alpha)$ 
 is driven by the shape of $f(x)$ while
 their $y$ or $\alpha$-profile is analogous to the
 shape of two-body distribution amplitudes like
 $[y(1-x-y)]^n$, \mbox{$[(1-x)^2 - \alpha^2]^n$.}  
 Fixing the profile parameter $n$  gives 
 simple  models \cite{ddee,sssdd}
for DDs which can be converted into models for SPDs.

In the present paper,
our main goal  is to study
the self-consistency of these models with respect to
the pQCD evolution. 
In Section II, we briefly review the basic 
elements of the  formalism of
  double distributions, discuss
their support and   symmetry properties and relation
to usual parton densities. In Section III, 
we describe  factorized profile models for 
double distributions and give explicit model
expressions for  skewed 
distributions. In Section IV we consider a 
practically important case when skewedness parameters
$\zeta$ or $\xi$ are small. 
 The factorized models for DDs in this case 
 can be taken in a very  simple form
 $\tilde f (x,\alpha; t=0) = f(x) \rho (\alpha)$,
 where $ \rho (\alpha)$ is a normalized profile function.
 As a result, SPDs $\tilde {\cal F}_{\zeta}(X)$ 
 (or $H( \tilde x, \xi)$) in this model
 are obtained    by  averaging the 
 relevant forward distribution $\tilde f(x)$
 over the interval $(X- \zeta,X)$ [or $(\tilde x -  \xi, \tilde x +  \xi)$] 
 with the weight $\rho (\alpha)$ 
 (we use the convention \cite{sssdd} that
 ``tilded'' parton distributions are those defined 
 on the $(-1,1)$ interval).  
 In Section V, we  study the impact
 of the pQCD evolution on the profile function $\rho (\alpha)$.
 Since  the $\alpha$-DDs are hybrids which look  like   usual parton densities
 wrt $x$ and like distribution amplitudes 
 wrt  $\alpha$, the  simplest  renormalization 
 properties at one loop have the combined 
 $x^m C_l^{3/2+m} (\alpha)$ moments of   $\tilde f (x,\alpha)$.
 As a result,  independently 
of the initial condition,  
 the $\alpha$-profile
 of 
 the $x^n$ moment $\tilde f_n (\alpha)$ 
 of the $\alpha$-DD  $\tilde f (x,\alpha)$ under the pQCD evolution 
asymptotically 
 tends to $(1-\alpha^2)^{n+1}$.  
 We investigate the ``asymptotic profile model''
 in which  $\tilde f_n
(\alpha)$  are given by their asymptotic form and show that
it imposes  a remarkable correlation $\tilde f (x,\alpha) =
F(x/(1-\alpha^2))$ between the   $x$-dependence of 
the $\alpha$-DDs  and their 
$\alpha$-profile. 
To study the impact of pQCD evolution on 
the DD based  models  of SPDs, we 
perform an explicit numerical evolution 
of SPDs.  In Section VI, we describe a simple 
algorithm for the  leading-log evolution of 
SPDs based on  direct iterative
convolutions of evolution kernels $W_{\zeta} (X,Z)$  
with SPDs ${\cal F}_{\zeta} (Z)$. 
In section VII, we  discuss the results 
of our numerical  calculations.  
In Appendix A, we show that  the 
approximation  (used  in Ref. \cite{sgmr}) 
in which the Gegenbauer moments 
of SPDs do not depend on skewedness,
is equivalent to the  asymptotic profile model 
for DDs.  In Appendix B, we present 
explicit form of evolution equations
for SPDs used in our numerical calculations.

\section{Double distributions}

In the pQCD factorization treatment of hard
electroproduction processes, the 
nonperturbative information is accumulated in the
nonforward matrix elements
$\langle p-r \,  |     \,  {\cal O} (0,z)  \,  |     \,  p \rangle $
of   light cone operators $ {\cal O} (0,z) $. 
For  $z^2=0$ the matrix elements 
depend on the relative coordinate $z$
through two Lorentz invariant variables $(pz)$ and $(rz)$.
In the forward case, when $r=0$, 
one obtains the usual quark helicity-averaged densities   
by  Fourier transforming   the relevant  matrix element 
with respect to $(pz)$
\begin{equation} 
\langle p,s'\, \,  |     \,  \, \bar \psi_a(0) \hat z 
E(0,z;A)  \psi_a(z) \, \,  |     \,  \, p,s \rangle \,  |     \, _{z^2=0} 
 =  \bar u(p,s')  \hat z u(p,s)  
   \int_0^1  \,  
 \left ( e^{-ix(pz)}f_a(x) 
  -   e^{ix(pz)}f_{\bar a}(x)
\right ) \, dx \, , 
\label{33} \end{equation} 
where $E(0,z;A)$ is the gauge link,  
 $\bar u(p',s'), u(p,s)$ are the Dirac
spinors and we use the  notation
$\gamma_{\alpha} z^{\alpha} \equiv \hat z$.
The  functions $ f_a(x) $, $ f_{\bar a}(x)$ can be also treated 
as  components of the ``tilded'' distribution
$ \tilde f_a(x) = f_a(x) \theta (x>0) - f_{\bar a}(x)\theta (x<0)$,
whose support extends to $[-1 \leq x \leq 1]$.
In the nonforward case,  we can  
use the  double Fourier  representation 
with respect to both $(pz)$ and $(rz)$ \cite{npd}: 
\begin{eqnarray} 
&& \langle p',s'\, \,  |     \,  \, \bar \psi_a(0) \hat z 
E(0,z;A)  \psi_a(z) \, \,  |     \,  \, p,s \rangle \,  | 
    \, _{z^2=0} 
\label{31}  \\ &&  =  \bar u(p',s')  \hat z u(p,s)  
\int_0^1  dy   \int_{-1}^1  \,  
  e^{-ix(pz)-iy(r z)} \, \tilde F_a(x,y;t) \, 
 \theta( 0 \leq x+y \leq 1) \, dx  \, + \,  ``O(r){\rm -terms}"  \, , 
\nonumber 
 \end{eqnarray} 
where  the
``$O(r)$-terms'' stands for  contributions  
which have the structure   $\bar u(p',s')  r^{\mu} z^{\nu} 
\sigma_{\mu \nu} u(p,s)\Phi((pz),(rz))$ \cite{ji}  and  
$(rz) \bar u(p',s')    u(p,s) \Psi ((rz))$  
\cite{jirev,poweiss} and  vanish 
in the $r \to 0$ limit.

\begin{figure}[htb]
\mbox{
   \epsfxsize=13.5cm
 \epsfysize=7.5cm
 \hspace{1.5cm}  
  \epsffile{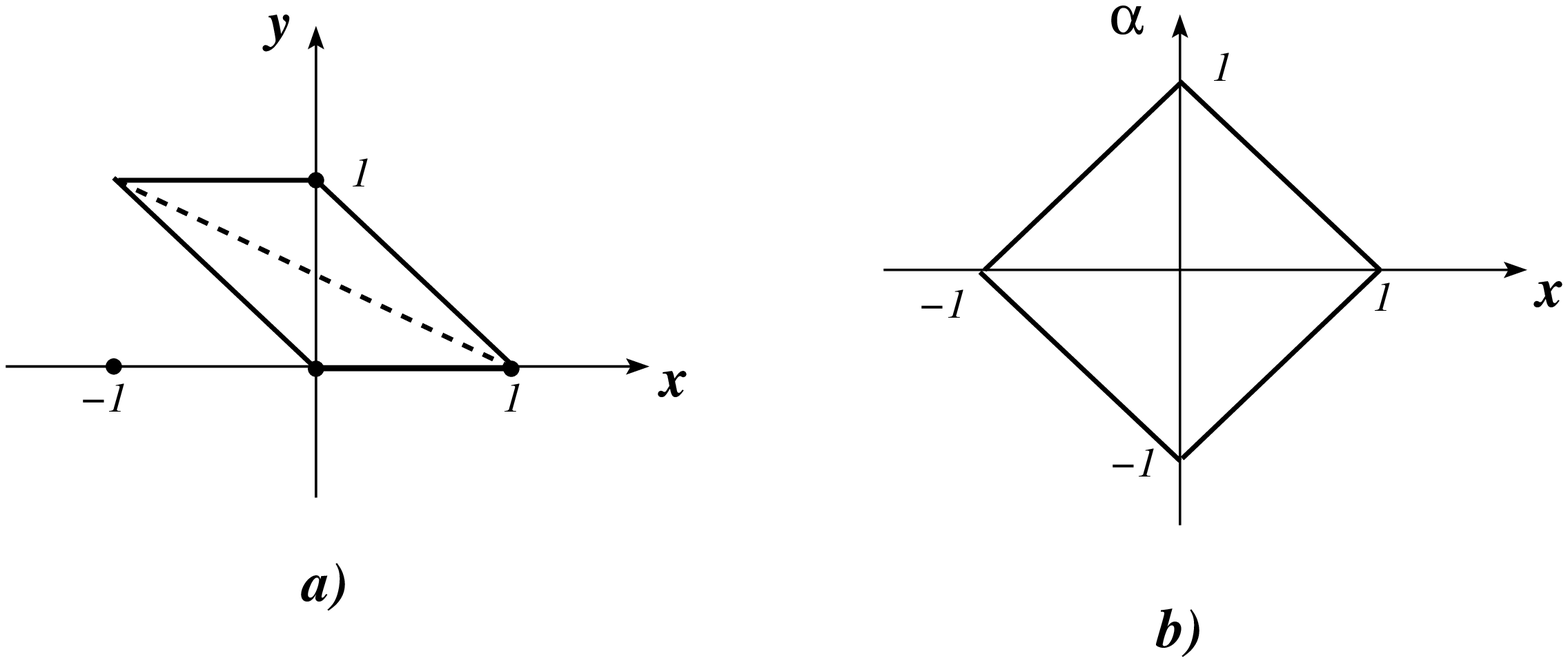}  }
  \vspace{-1.5cm} 
{\caption{\label{fg:support} $a)$ Support region 
and symmetry line  $y = \bar x/2$ for 
$y$-DDs 
$\tilde F(x,y;t)$. $b)$ Support region 
for $\alpha$-DDs $\tilde f (x, \alpha)$. 
   }}
\end{figure}

 For any Feynman 
diagram, the spectral constraints 
$-1 \leq x \leq 1$, $0 \leq y \leq 1$, $0 \leq x+y
\leq 1$ were  proved in the $\alpha$-representation
\cite{npd} using the approach of Ref.
\cite{spectral}. The support area 
for  the {\it double distribution} $ \tilde F_a(x,y;t)$ 
is shown on Fig.\ref{fg:support}$a$. 

\vspace{-1cm}

Taking the $r =0$ limit of Eq.  (\ref{31}),
one obtains    
``reduction formulas'' 
relating   the double distribution $\tilde F_{a}(x,y;t=0)$  
to the quark and antiquark parton  densities
\begin{equation} 
\int_0^{1-x} \, \tilde F_{a}(x,y;t=0)|_{x>0} \, dy= 
 f_{a}(x) \hspace{1cm} ; \hspace{1cm} 
\int_{-x}^1 \, \tilde F_{ a}(x,y;t=0)|_{x< 0}\, dy= 
-  f_{\bar a}(-x) \,  \label{eq:redfsym} \, .
 \end{equation}
Hence, the  positive-$x$
and negative-$x$  components of the double 
distribution $\tilde F_{a}(x,y;t) $
can be treated as nonforward generalizations of 
quark and antiquark densities, respectively.
If we define the ``untilded'' DDs
\begin{equation} 
F_{a}(x,y;t) = \tilde F_{a}(x,y;t)|_{x>0}  
\hspace{1cm} ; \hspace{1cm} 
F_{\bar a}(x,y;t) = - \tilde F_{a}(-x,1-y;t)|_{x<0}   \, ,
\label{abara}   \end{equation}
then $x$ is always positive and 
 the reduction formulas  have  the same form
\begin{equation} 
\int_0^{1-x} \, F_{a,\bar a}(x,y;t=0)|_{x \neq 0} \, dy= 
 f_{a,\bar a}(x) 
 \label{34} \end{equation}
in both cases. The new distributions both ``live'' on the triangle
$0 \leq x,y \leq 1, \, 0 \leq x +y \leq 1$.
Taking $z$ in the lightcone ``minus''  direction,
we arrive at  the   parton interpretation 
of  functions $ F_{a, \bar a} (x,y;t )$ as probability amplitudes for an 
outgoing  parton to carry the fractions $xp^+$
and $yr^+$ of the external 
momenta $r$ and $p$. 
The  double distributions 
$F(x,y;t)$  are universal functions 
describing the flux of $p^+$ and $r^+$ 
independently of the ratio $r^+/p^+$.

\begin{figure}[htb]
\mbox{
   \epsfxsize=12cm
 \epsfysize=5cm
 \hspace{0.5cm}  \epsffile{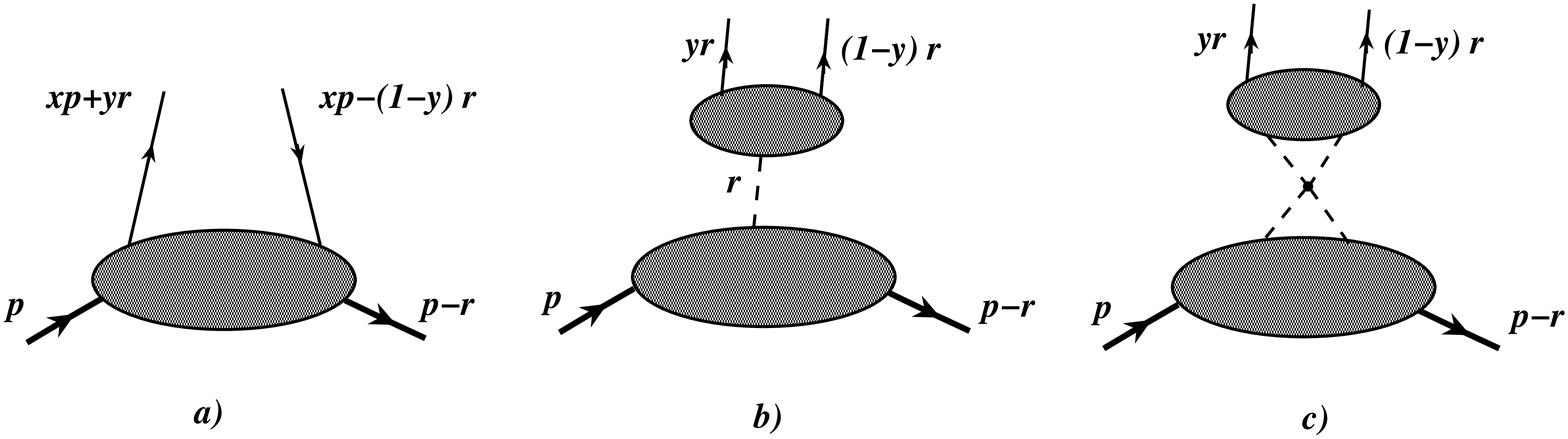} 
 \hspace{-1cm}  \epsfxsize=5cm
 \epsfysize=5cm \epsffile{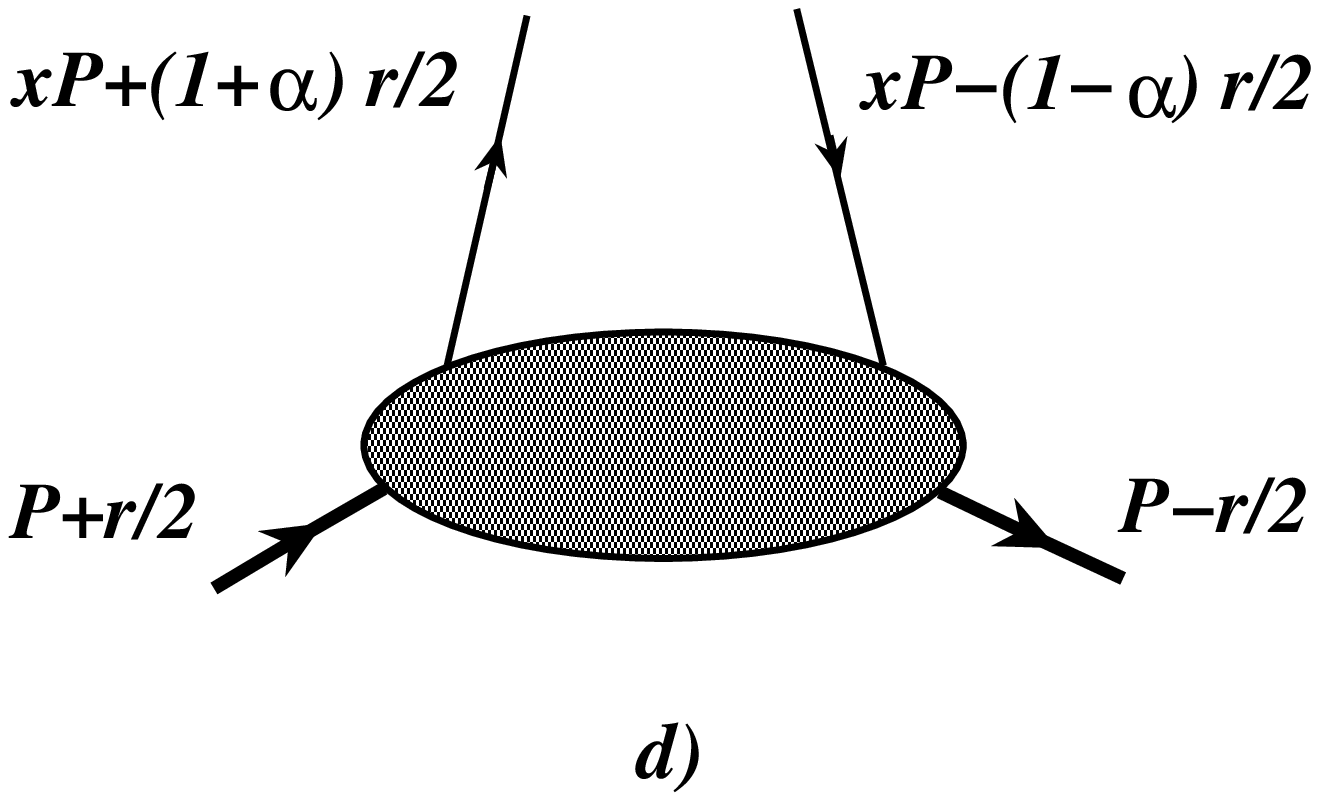} }
  \vspace{-1.5cm}
{\caption{\label{fg:exchange} $a)$ Parton picture
in terms of $y$-DDs; $b)$   meson-like contribution;
$c)$   Polyakov-Weiss contribution; 
$d)$ parton picture
in terms of $\alpha$-DDs.
   }}
\end{figure}

The   functions 
$\tilde F (x,y;t)$  may   have  singular terms at $x=0$ 
proportional 
to $\delta (x)$ or its derivative(s).
In such terms the sensitivity
of the parton distribution to the plus-component
of the initial hadron momentum is lost,
and they  
have no projection  onto the usual parton densities.
We will denote them by $F_M (x,y;t)$ $-$  they may 
be interpreted as coming from the $t$-channel 
meson-exchange type contributions (see Fig.\ref{fg:exchange}b).  
As shown by Polyakov and Weiss \cite{poweiss},
the terms with quartic pion vertex (Fig.\ref{fg:exchange}c), in 
which the dependence on $(pz)$ is also lost, 
correspond to an 
independent   $(rz) \bar u(p',s')    u(p,s) \Psi ((rz))$ 
type contribution. Both the meson-exchange and 
Polyakov-Weiss terms are invisible 
in the forward limit, hence the existing 
knowledge of the usual parton densities 
cannot be used  to   constrain
these terms. In what follows,
we consider only the ``forward visible parts''
of SPDs which are obtained by scanning 
the $x \neq 0$ parts of the relevant DDs.

To make the description more symmetric 
with respect to the  initial and final hadron momenta,  
we can  treat nonforward matrix elements as 
functions of $(Pz)$ and $(rz)$, where $P=(p+p')/2$
is the average hadron momentum. 
The relevant  double distributions 
$\tilde f_a(x,\alpha\,;t)$ [which we 
will call $\alpha$-DDs to distinguish them
from $y$-DDs $F(x,y;t)$] are defined by \cite{realco,sssdd}   
\begin{eqnarray}
\langle p' |  \bar \psi_a (-z/2) \hat z
\psi_a (z/2) |p \rangle \, = 
\bar u(p') \hat z u(p) \, \int_{-1}^1 \, dx \int_{-1+|x|}^{1-|x|} 
 e^{-ix(Pz)-i\alpha (rz)/2}  \tilde f_a(x,\alpha;t) 
 \,
d\alpha  \,   
 + ``O(r){\rm -terms}"\, .
 \label{17} \end{eqnarray}
The support area for $\tilde f_a(x,\alpha;t)$ is shown 
in Fig.\ref{fg:support}b.
Again, the usual forward densities 
$f_a(x)$ and $f_{\bar a} (x)$ 
are given by integrating 
$\tilde f_a(x,\alpha\,;\,t=0)$  over vertical lines
$x = {\rm const} $ for $x>0$ and $x<0$, respectively.
 Due  to hermiticity
and time-reversal invariance properties of
nonforward matrix elements,  the $\alpha$-DDs
are even functions of $\alpha$: 
$$
\tilde f_a(x,\alpha;t) = \tilde f_a(x, - \alpha;t) \, . 
$$
For our original $y$-DDs $F_{a, \bar a} (x,y;t)$, this 
corresponds to the ``Munich'' symmetry 
with respect to  the interchange 
$y \leftrightarrow 1-x-y$ established in Ref. \cite{lech}. 
The  $a$-quark contribution 
into the 
 flavor-singlet 
operator
can be  parametrized either by  $y$-DDs $\tilde F_a^S (x,y;t)$ or  
by $\alpha$-DDs $\tilde f_a^S(x, \alpha \, ; \, t)$.
The latter 
are even functions of   $\alpha$ and odd functions of $x$:
\begin{equation}
  \tilde f^{S}_a (x,\alpha ;t)   =   
  \{ f_a(|x|,|\alpha | ;t) + f_{\bar a} (|x|,|\alpha| ;t) \} \, 
  {\rm sign} (x) +f_M^S(x,\alpha ;t) \ .
\label{singlet} \end{equation} 
The valence 
 quark functions $\tilde f_a^{V}(x, \alpha \, ; \, t)$
are even functions of both $\alpha$  and $x$:
\begin{equation}
  \tilde f^{V}_a (x,\alpha ;t)   =   
  f_a(|x|,|\alpha| ;t) - f_{\bar a} (|x|,|\alpha|;t) + f_M^V(x,\alpha ;t) \ .
\label{valence} \end{equation}

It is convenient to define the  gluonic 
 $\alpha$-DD $\tilde f^G(x,\alpha;t)$  
in such a way that its integral over $\alpha$ for $t=0$, 
 also gives  the usual forward
gluon density $\tilde f^G(x)$: 
\begin{eqnarray} 
&& \langle P-r/2  \,  |      \,   
z_{\mu}  z_{\nu} G_{\mu \alpha}^a (-z/2) E_{ab}(-z/2,z/2;A) 
G_{ \alpha \nu}^b (z/2)\,   |       \,P+r/2  \rangle \,  |     \, _{z^2=0}
  \label{36}  \\  &&  
= \bar u(p')  \hat z 
 u(p) \, (Pz) \int_0^1 dx   \int_{-1 + |x|}^{1-|x|}  \, 
  e^{-ix(Pz)-i\alpha (r z)/2} 
 \, x \tilde f^G(x,\alpha;t)  
 \,  d\alpha  + O(r) -{\rm terms} \,  .
\nonumber
 \end{eqnarray} 
 The gluon SPD $H^G ( \tilde x, \xi; \mu)$ is constructed in this case
 from  $x \tilde f^G (x, \alpha ;t )$.  
Just like the singlet quark distribution,
the function  $\tilde f^G (x,\alpha;t)$  is an odd function 
of $x$.

\section{Models for double and skewed distributions}

The  reduction formulas and   interpretation of
the  $x$-variable  as the  fraction of 
  $p$ (or $P$) momentum 
suggest that the   profile of $F(x,y)$ (or $f(x,\alpha)$)  
in  $x$-direction is basically determined by the shape 
of $f(x)$. 
On the other hand, the profile in  $y$ (or $\alpha$) direction  
characterizes the spread of the parton momentum induced by
the momentum transfer $r$. 
In particular, since 
the $\alpha$-DDs $\tilde f(x,\alpha)$ 
are even functions of $\alpha$,
it make sense to write 
\begin{equation}
\tilde f(x,\alpha) =  h(x,\alpha) \,  \tilde f(x)  \, ,  \label{65n}
 \end{equation}
 where $h(x,\alpha)$ is an even function of $\alpha$ 
 normalized by 
\begin{equation}
 \int_{-1+|x|}^{1-|x|} h(x,\alpha) \, d\alpha \, =1.
 \end{equation}
We may expect that 
the $\alpha$-profile of $h(x,\alpha)$  
is similar to that of a symmetric distribution amplitude (DA) 
$\varphi (\alpha)$.  Since $|\alpha| \leq 1- |x| $, to get a 
more complete  analogy
with DA's, 
it makes sense to rescale $\alpha$ as $\alpha = (1-|x|)  \beta$
introducing  the  variable $\beta$ with $x$-independent limits:
$-1 \leq \beta \leq 1$. 
 The simplest model is to assume 
that the profile in the $\beta$-direction is  
 a  universal function  $g(\beta)$ for all $x$. 
Possible simple choices for  $g(\beta)$ may be  $\delta(\beta)$
(no spread in $\beta$-direction),  $\frac34(1-\beta^2)$
(characteristic shape for asymptotic limit 
of nonsinglet quark distribution amplitudes), 
 $\frac{15}{16}(1-\beta^2)^2$
(asymptotic shape of gluon distribution amplitudes), etc.
In the variables $x,\alpha $, this gives   
\begin{equation}
h^{(\infty)} (x,\alpha) =  \delta(\alpha) \,   \ , \
h^{(1)}(x,\alpha) = \frac{3}{4} 
\frac{ (1- |x|)^2 - \alpha^2}{(1-|x|)^3} \,   \ , \
h^{(2)}(x,\alpha) = \frac{15}{16} 
\frac{[(1- |x|)^2 - \alpha^2]^2}{(1-|x|)^5} \,   \  . \label{mod123} 
 \end{equation}
 These models can be treated as specific
 cases of the general profile function 
 \begin{equation}
 h^{(b)}(x,\alpha) = \frac{\Gamma (2b+2)}{2^{2b+1} \Gamma^2 (b+1)}
\frac{[(1- |x|)^2 - \alpha^2]^b}{(1-|x|)^{2b+1}} \,  , \label{modn} 
 \end{equation}
whose width is governed by the parameter $b$.

\begin{figure}[htb]
\mbox{
   \epsfxsize=14cm
 \epsfysize=8cm
 \hspace{1cm}  
  \epsffile{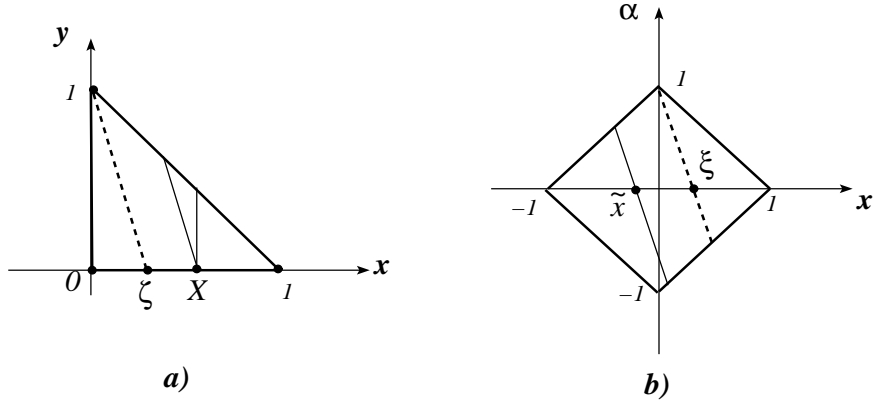}  }
  \vspace{-1cm}
{\caption{\label{fg:spds}    Integration lines 
for integrals relating SPDs and DDs. 
   }}
\end{figure}

The  coefficient of proportionality 
$\zeta = r^+/p^+$  (or $\xi = r^+/2P^+$)  between 
the plus components of the momentum transfer 
and initial (or  average) momentum specifies 
the {\it skewedness}  of the matrix elements.
The characteristic   feature  implied by
 representations for double distributions 
 [see, e.g., Eq.(\ref{31})]
 is the absence 
of the $\zeta$-dependence in 
the DDs   $F(x,y)$ and $\xi$-dependence in $f(x,\alpha)$.
An alternative way to parametrize 
nonforward matrix elements of light-cone operators
is to use   
 $\zeta$  (or $\xi$) and the {\it total } 
 momentum fractions  $X \equiv x+y \zeta$ 
(or $\tilde x \equiv  x +\xi \alpha$) 
  as  independent 
variables. 
Integrating each particular  double   distribution 
over $y$ gives  the nonforward  parton distributions  
\begin{eqnarray} && 
{\cal F}_{\zeta}^{i} (X) = \int_0^1 dx \int_0^{1-x} 
\, \delta (x+\zeta y -X) \, F_i(x,y) \, dy \nonumber \\ &&
=
\theta(X \geq \zeta) 
 \int_0^{ \bar X / \bar \zeta } F_{i}(X-y \zeta,y) \, dy + 
 \theta(X \leq \zeta)  \int_0^{ X/\zeta} F_{i}
 (X-y \zeta,y) \, dy \,  , 
\label{71}  \end{eqnarray}
where $\bar \zeta \equiv 1- \zeta$.
The two components of NFPDs correspond to positive
($X> \zeta$) and negative ($X< \zeta$)
values of the fraction $X' \equiv X - \zeta$ 
associated with the ``returning'' parton.
As explained in refs. \cite{compton,npd},
the second component  can be interpreted as the
probability amplitude for the initial hadron with momentum 
$p$ to split into the final hadron with momentum $(1-\zeta)p$
and the two-parton state with total momentum $r=\zeta p$
shared by the partons
in fractions $Yr$ and $(1-Y)r$, where $Y=X/\zeta$.

The relation between ``untilded'' NFPDs
and  DDs  can be     illustrated
on the ``DD-life triangle'' 
defined by $0 \leq x,y,x+y \leq 1$ (see Fig.\ref{fg:spds}a).
Specifically, to get ${\cal F}_{\zeta} (X)$,  one should 
integrate $F(x,y)$ over $y$ along a straight
line  $x=X- \zeta y$. Fixing some value of  $\zeta$,
one deals with  a set of parallel lines  intersecting  the $x$-axis
at $x=X$. The upper limit of the $y$-integration
is determined by intersection of this line
either with the line $x+y=1$ (this happens if 
$X > \zeta$)  or with the $y$-axis (if $X < \zeta$).
The line corresponding to $X=\zeta$
separates the triangle into two parts
generating the  two components of the
nonforward parton distribution.

In a similar way, we can write the relation 
between OFPDs 
$H(\tilde x,\xi;t)$  
 and the 
$\alpha$-DDs $\tilde f(x,\alpha;t)$ 
\begin{equation} 
  H (\tilde x,\xi; t)=  \int_{-1}^1 dx \int_{-1+|x|}^{1-|x|}  
\, \delta (x+ \xi \alpha - \tilde x) \, 
\tilde f (x,\alpha;t) \, d\alpha  \, . \label{710} 
 \end{equation}
It should noted that  
 OFPDs as defined by 
X. Ji \cite{ji} correspond to parametrization
of the nonforward matrix element by a Fourier integral 
with a single  common exponential with $-1 \leq \tilde x \leq 1$,   
i.e., OFPDs are equivalent to   ``tilded'' NFPDs. 
The  delta-function in Eq.(\ref{710}) specifies 
the line of  integration in the
$\{ x, \alpha \}$ plane. For definiteness,  we will assume below
that 
$\xi$ is positive.

Information contained in SPDs originates from two  
physically different sources: meson-exchange type contributions  
${\cal F}_{\zeta}^M(X)$ 
coming from the singular $x=0$ parts of DDs
and  the functions 
${\cal F}_{\zeta}^a(X)$,  ${\cal F}_{\zeta}^{\bar a}(X)$
 obtained by scanning the $x \neq 0$  parts of  
DDs $F^a(x,y)$, $F^{\bar a} (x,y)$. 
The support of exchange contributions is restricted 
to $0 \leq X \leq \zeta$. Up to rescaling, the function
${\cal F}_{\zeta}^M(X)$  has the same shape for all $\zeta$.
For any nonvanishing $X$, these exchange terms  become  invisible 
in the forward limit $\zeta \to 0$.  
On the other hand, the support of functions 
${\cal F}_{\zeta}^a(X)$,  ${\cal F}_{\zeta}^{\bar a}(X)$
in general covers the whole   $0\leq X \leq 1$ region. 
Furthermore, the forward limit of such  SPDs as 
${\cal F}_{\zeta}^{a, \bar a}(X)$ is   
rather well known  from inclusive measurements.  
Hence, information  contained in the usual (forward) densities
$f^a(x)$, $f^{\bar a}(x)$ can be used to 
restrict the models for 
${\cal F}_{\zeta}^a(X)$,  ${\cal F}_{\zeta}^{\bar a}(X)$.

Let us consider 
SPDs  constructed using    simple 
models of DDs specified in  Section III.
In particular,  the model
$f^{(\infty )}(x,\alpha) = \delta (\alpha)
f(x)$ (equivalent to $F^{(\infty)}(x,y) = \delta (y -\bar x/2)
f(x)$), gives the simplest    model 
  $H^{(\infty)}(\tilde x,\xi; t=0) = f(x)$ in which  OFPDs at $t=0$
   have no $\xi$-dependence.
For NFPDs this gives  
\begin{equation}
{\cal F}_{\zeta}^{(\infty)} (X) = \frac{\theta(X \geq \zeta/2)}{1-\zeta/2}
 f \left (\frac{X-\zeta/2}{1-\zeta/2} \right ) \, ,
 \label{model}
  \end{equation}
i.e.,  NFPDs for non-zero  $\zeta$ are obtained from 
the forward distribution $f(X)\equiv {\cal F}_{\zeta=0} (X)$  
 by   shift and rescaling.

\begin{figure}[htb]
\mbox{
   \epsfxsize=8cm
 \epsfysize=5cm
 \hspace{0cm}  
  \epsffile{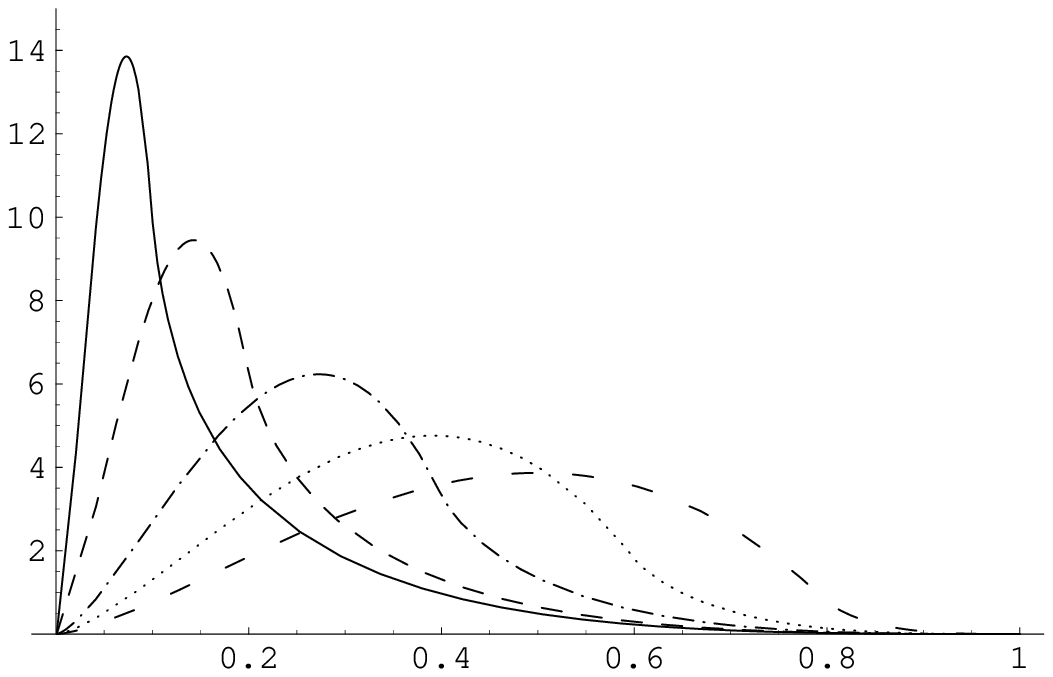}  } \hspace{0cm}
\mbox{
   \epsfxsize=8cm
 \epsfysize=5cm
 \hspace{0cm}  
  \epsffile{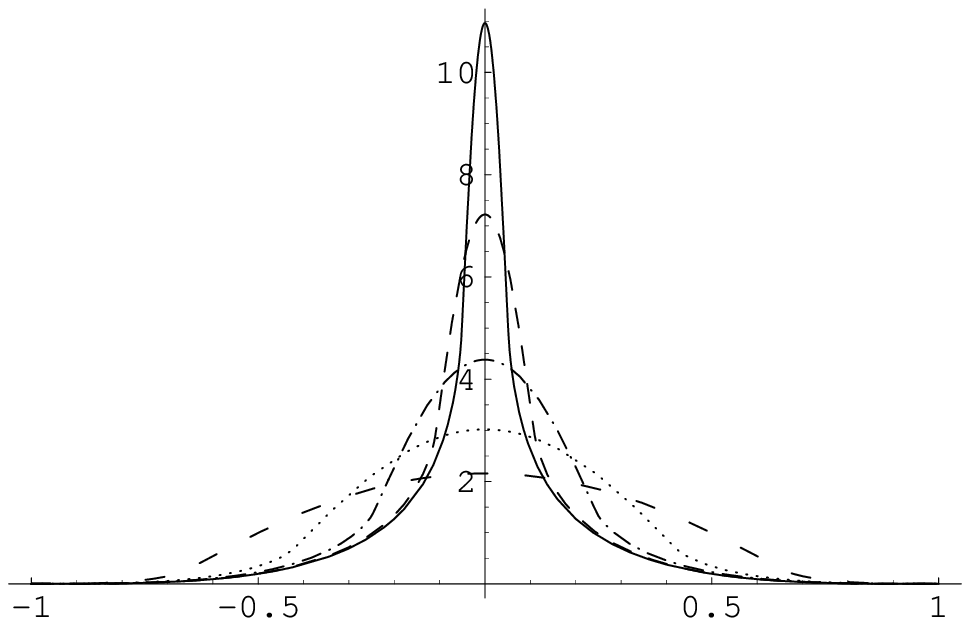}  }
{\caption{\label{q-at-diff-z}
Valence quark  distributions: untilded NFPDs $F^{q}_\zeta(x)$ (left)
and OFPDs $H ^1_V(\tilde x,\xi)$ (right) with $a= 0.5$ 
for several values
of $\zeta$, namely,  0.1, 0.2, 0.4, 0.6, 0.8 and corresponding values
of $\xi=\zeta / (2-\zeta)$. Lower curves
correspond to larger values of $\zeta$.}}
\end{figure}

In case of  the $b=1$ and $b=2$  models, simple  analytic results 
can be obtained only for some   explicit forms of  $f(x)$.
For the ``valence quark''-oriented ansatz $\tilde f^{(1)}(x,\alpha)$,
the following choice of a normalized distribution
\begin{equation}  f^{(1)}(x) = 
 \frac{\Gamma(5-a)}{6 \,  \Gamma(1-a)} \, 
  x^{-a} (1-x)^3  \label{74} \end{equation}
is   both  close 
to phenomenological   quark distributions
and   produces a simple expression
for the double distribution since the denominator
$(1-x)^3$ factor in Eq. (\ref{mod123}) is canceled.
As a result, the integral in Eq. (\ref{710})
is easily performed and   we get
\begin{equation}
H^{1 }_V(\tilde x, \xi)|_{|\tilde x| \geq \xi}  = \frac1{\xi^3} 
 \left ( 1- \frac{a}{4} \right ) 
\left \{ \left[ (2-a) \xi (1- \tilde x) (x_1^{2-a} + x_2^{2-a}) +
(\xi^2 -\tilde x)(x_1^{2-a} - x_2^{2-a}) \right ] \, \theta (\tilde x) 
+ ( \tilde x \to -\tilde x) \right \}  \label{outs} 
\end{equation}
for  $|\tilde x |\geq \xi$ and 
\begin{equation}
H^{1 }_V (\tilde x, \xi)|_{|\tilde x| \leq \xi}  = \frac1{\xi^3} 
\left ( 1- \frac{a}{4} \right ) \left \{ x_1^{2-a}[(2-a) \xi (1- \tilde x) +
(\xi^2 -x)] + ( \tilde x \to -\tilde x) \right \} \label{middles}
\end{equation}
in the middle $ -\xi \leq \tilde x \leq \xi$ region.
We use here  the notation $x_1=(\tilde x + \xi)/(1+\xi)$
  and 
$x_2=(\tilde x - \xi)/(1-\xi)$ \cite{jirev}.
To extend these expressions onto negative values of 
$\xi$, one should  substitute $\xi$ by $|\xi|$.
One can check, however, that no odd powers of $|\xi|$ 
would appear in the $\tilde x^N$ 
moments of $H^{1V }(\tilde x, \xi)$.
Furthermore, these expressions are explicitly non-analytic 
for $x = \pm \xi$. 
This  is true even if $a$ is integer.
Discontinuity at $x = \pm \xi$, however, appears only 
in the second derivative
of $H^{1V }(\tilde x, \xi)$, 
i.e., the model curves for $H^{1V }(\tilde x, \xi)$
look very smooth (see Fig.\ref{q-at-diff-z}).
The explicit expressions for NFPDs in this 
model were given in ref.\cite{ddee}. The relevant curves 
are also shown in Fig.\ref{q-at-diff-z}.

For   $a=0$, the 
$x>\xi$ part of OFPD has the same $x$-dependence 
as its forward limit, differing from it by an overall $\xi$-dependent 
factor only:
\begin{equation}
H^{1V }(\tilde x, \xi)|_{a=0} = 
4 \, \frac{(1-|\tilde x|)^3}{(1-\xi^2)^2} 
\, \theta (|\tilde x| \geq \xi) \, 
+ 2\, \frac{\xi +2 -3 \tilde x^2/\xi}{(1+\xi)^2} \, 
 \theta (|\tilde x| \leq \xi)
\, .    \label{(1-x)^3}
\end{equation} 
The  $(1-|\tilde x|)^3$ behaviour can be 
trivially continued into the $|\tilde x| < \xi$ 
region. However, the actual behaviour
of $H^{1V }(\tilde x, \xi)|_{a=0}$ in this region 
is given by a different function.
In other words, $ H^{1V }(\tilde x, \xi)|_{a=0}$
can be represented as a sum of a function analytic at 
border points and a contribution  whose support 
is restricted by $|\tilde x| \leq \xi$.  
It should be emphasized that despite its DA-like 
appearance, this contribution 
should not be treated as an exchange-type term. 
It is generated by regular $x \neq 0$ part of DD,
and, unlike $\varphi (\tilde x / \xi)/\xi$ functions
 changes its shape with $\xi$ becoming very small  for small $\xi$.

\begin{figure}[htb]
\mbox{
   \epsfxsize=8cm
 \epsfysize=5cm
 \hspace{0cm}  
  \epsffile{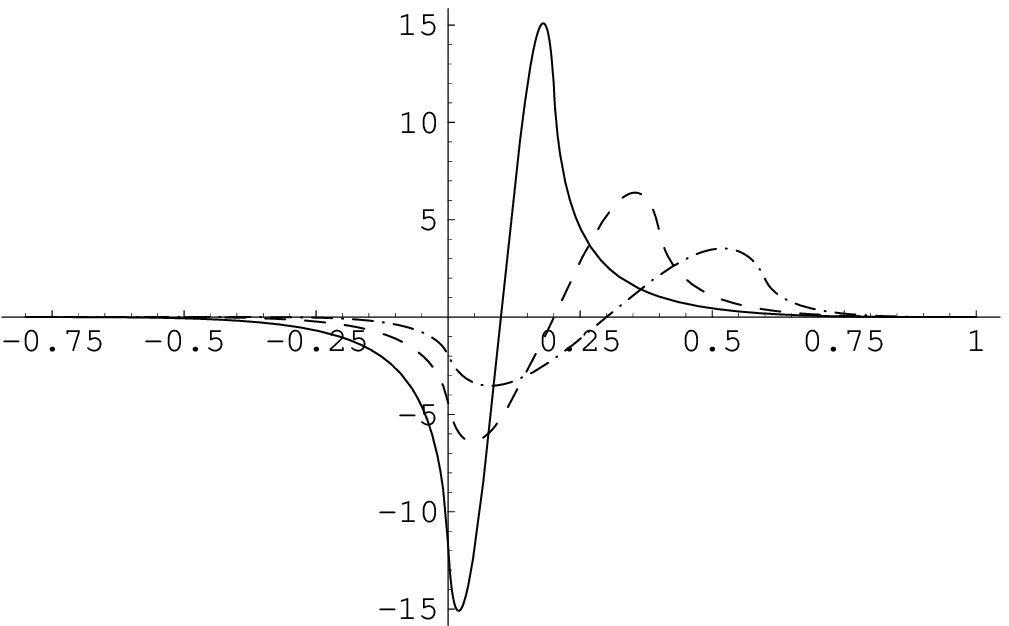}  } \hspace{0cm}
\mbox{
   \epsfxsize=8cm
 \epsfysize=5cm
 \hspace{0cm}  
  \epsffile{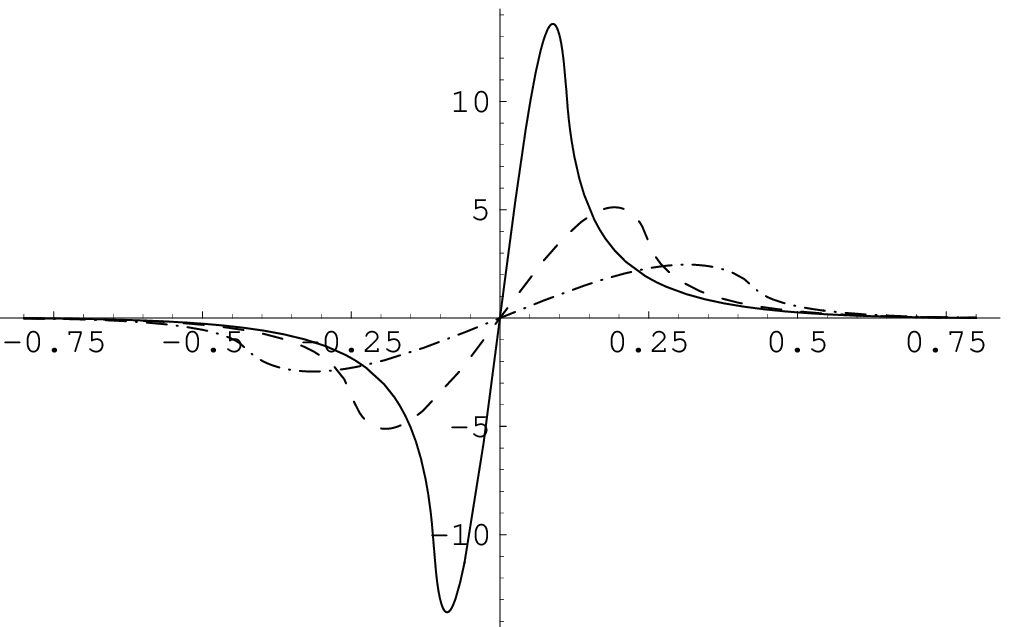}  }
{\caption{\label{qs-at-diff-z}
Singlet quark distributions: tilded NFPDs $\tilde F^{S}_\zeta(x)$ (left)
and  OFPDs $H ^1_S(\tilde x,\xi)$ (right) for 
several $\zeta$ values
 0.2, 0.4, 0.6 and corresponding values
of $\xi=\zeta / (2-\zeta)$. Lower curves
correspond to larger values of $\zeta$. Forward distribution
is modeled by $(1-x)^3/x$.}}
\end{figure}

For  the singlet quark distribution, the  $\alpha$-DDs
$\tilde f^S( x, \alpha)$ should be odd functions 
of $x$. Still, we can use  the model like (\ref{74}) for 
the $x>0$ part, but 
take $\tilde f^S( x, \alpha)|_{x \neq 0} 
= A \, f^{(1)}( |x|, \alpha)\, {\rm sign} (x)$. 
Note, that the integral (\ref{710})  producing $H^S(\tilde x, \xi)$ 
in the $|\tilde x| \leq \xi$ region
would diverge  for  $\alpha \to \tilde x /\xi$  
 if $a \geq  1$, which is the usual case 
 for standard parametrizations of singlet quark 
 distributions for sufficiently large $Q^2$. 
However, due to the antisymmetry of  $\tilde f^S( x, \alpha)$
wrt  $x \to -x$ and its symmetry wrt $\alpha \to -\alpha$,
the singularity at 
$\alpha = \tilde x /\xi$ can be  integrated  
using the principal value 
prescription   which in this case 
produces the $x\to -x$ antisymmetric version 
of Eqs.(\ref{outs}) and (\ref{middles}). For $a=0$, 
its middle part  reduces to 
\begin{equation}
H^{1S}(\tilde x, \xi)|_{|\tilde x| \leq \xi, a=0} = 
 2x\, \frac{3 \xi^2 -2 x^2 \xi - x^2}{\xi^3 (1+\xi)^2} 
  \,  .
\end{equation}
 The shape of singlet SPDs in this model is shown in Fig.
\ref{qs-at-diff-z}

\section{SPDs at small skewedness} 
 
 To study  the deviation of skewed distributions  
 from their forward counterparts for small $\xi$ (or $\zeta$), 
let us consider 
 the 
$x \geq \xi$ part of $H(x,\xi)$
[see   Eq.(\ref{710})]  and use its expansion  in powers of $\xi$
\cite{ddee}:
\begin{equation}
H(\tilde x;\xi)|_{\tilde x \geq \xi} 
 = f(\tilde x) +  \xi^2 \left [ 
\frac12 \int_{- (1- \tilde x)}^{(1- \tilde x)}
\frac{ \partial^{2}f (\tilde x,  \alpha  )}{\partial \tilde x^{2}}
 \,\alpha  ^{2} \, d  \alpha   \,  + (1- \tilde x)^2
\left. \left (\frac{ \partial f (\tilde x,  \alpha  )}
 {\partial \alpha } 
 -2 \frac{ \partial f (\tilde x,  \alpha  )}
 {\partial \tilde x  } \right ) \right |_{\alpha = 1 - \tilde x}
 \right ] +  \ldots \, .
\label{exxi2} 
\end{equation} 
where  $f(\tilde x)$ is  the forward distribution.
For small $\xi$, the corrections 
are formally $O(\xi^2)$.
However,   if $f(x,\alpha)$  
has a singular 
behavior like $x^{-a}$, then 
$$\frac{ \partial^{2} f(\tilde x,  \alpha  )}{\partial \tilde x^{2}}
\sim \frac{a (1+a)}{\tilde x^2} f(\tilde x,  \alpha  )\ , $$
and the relative suppression of the first correction 
is $O(\xi^2/\tilde x^2)$.   Though the corrections 
are tiny for  $\tilde x \gg \xi $, in  the region  $\tilde x \sim
\xi$  it  has   no parametric smallness.
 It is easy to write expicitly all  the   terms 
 which are not suppressed in the $\tilde x \sim \xi \to 0$ limit
  \begin{equation}
H(\tilde x;\xi) = \sum_{k=0} 
\frac{\xi^{2k}}{(2k)!}   
 \int_{- 1}^{1}
\frac{ \partial^{2k} f (\tilde x,  \alpha  )}{\partial \tilde x^{2k}}
 \,\alpha  ^{2k} \, d  \alpha   \, + \ldots = 
 \int_{- 1}^{1} 
\tilde  f (\tilde x -\xi   \alpha, \alpha  )  \, d \alpha
  + \ldots \, 
  \, ,
 \end{equation}
 where the ellipses denote the terms vanishing in this  
 limit. This result can be directly obtained 
 from Eq. (\ref{710}) by noting that 
 for small $x$, we can neglect the $x$-dependence in  the 
 limits $\pm (1-|x|)$ of the $\alpha$-integration. 
 Furthermore, for small $x$ one  can also 
 neglect the $x$-dependence of the  profile
 function $h(x,\alpha)$ in Eq. (\ref{65n})
 and  take the model $\tilde f(x,\alpha) = \tilde f(x) \rho (\alpha)$
 with $\rho (\alpha)$ being a symmetric normalized weight 
 function on $-1 \leq \alpha \leq 1$. Hence, in the region where both
  $\tilde x$ and $\xi$ are small, we can approximate Eq. (\ref{710}) by 
 \begin{equation}
H(\tilde x;\xi) = {\rm ``P"}    
 \int_{- 1}^{1} 
\tilde  f (\tilde x -\xi   \alpha  ) \rho (\alpha) \, d \alpha
  + \ldots \, , \label{smallxi} 
 \end{equation}
 i.e., the OFPD $H(\tilde x;\xi)$ is obtained 
 in this case by averaging  the  usual (forward) 
 parton density $f(x)$ over the region 
 $\tilde x - \xi \leq x \leq \tilde x+\xi$ 
 with the weight $\rho (\alpha)$. 
 The  principal value prescription 
 ``P'' is only necessary in the case of singular quark 
 singlet distributions which are odd in $x$.   
 In terms of NFPDs, the relation is 
  \begin{equation}
\tilde  {\cal F}_{\zeta} (X )  = {\rm ``P"}    
 \int_{- 1}^{1} 
\tilde   f (X- \zeta (1+   \alpha)/2  ) \rho (\alpha)
 \, d \alpha
 + \ldots \, , \label{smallzeta}
 \end{equation}
 i.e., the average is taken over the region 
 $X -\zeta \leq x \leq X$.
 
 In fact,  for small values of the skewedness parameters
 $\xi, \zeta$, Eqs. (\ref{smallxi}), (\ref{smallzeta}) 
 can be used for all values of $\tilde x$ and $X$:
 if $\tilde x \gg \xi$, Eq. (\ref{smallxi}) gives the
 correct result 
 $H(\tilde x;\xi) = \tilde  f (\tilde x  ) + O(\xi^2)$.
 Hence, to get the ``forward visible'' parts of
 SPDs at small skewedness,
 one only needs to know the shape 
 of the normalized profile function $\rho (\alpha)$.

 The imaginary part of  hard exclusive 
 meson electroproduction amplitude is determined by
 the skewed distributions at 
 the border point. For this reason, the magnitude 
 of  ${\cal F}_{\zeta} (\zeta)$ [or $H(\xi, \xi)$]
   and its relation to the  forward densities
 $f(x)$   has a practical interest. 
 This example also gives a  possibility to study  
 the sensititivity of the results to the choice of the
 profile function.  
 Assuming   the infinitely narrow weight 
 $\rho(\alpha) = \delta (\alpha)$,
 we have ${\cal F}_{\zeta} (X ) = f(X-\zeta/2) + \ldots $ 
 and $H(x,\xi) = f(x)$.
 Hence, both ${\cal F}_{\zeta} (\zeta)$ and  $H(\xi, \xi)$
 are given by $f(x_{Bj}/2)$ since $\zeta = x_{Bj}$
 and $\xi =x_{Bj}/2 +\ldots$.  Since the argument
 of $f(x)$ is twice smaller than in deep inelastic scattering,
this results in an enhancement factor. In particular, if 
  $f(x) \sim x^{-a}$ for small  $x$, the ratio
  ${\cal R} (\zeta ) \equiv {\cal F}_{\zeta} (\zeta ) /f(\zeta )$ is 
  $2^a$. 
 The use of a wider profile  function $\rho (\alpha)$ produces further
 enhancement. For example, taking the normalized profile 
 function 
 \begin{equation}
 \rho_b (\alpha)\equiv  \frac{\Gamma (b+3/2) }{ \Gamma (1/2) 
 \Gamma (b+1) } (1- \alpha^2)^b = 
  \frac{\Gamma (2b+2)}{2^{2b+1} \Gamma^2 (b+1)}
 (1-\alpha^2)^b \label{rhon}
 \end{equation} 
 and
$f(x) \sim x^{-a}$ we get 
 \begin{equation}{\cal R}^{(b)} (\zeta) \equiv 
 \frac{{\cal F}_{\zeta}^{(b)}   (\zeta ) }{f(\zeta )}
  = \frac{\Gamma (2b+2)\Gamma (b-a +1)}  
  {\Gamma (2b-a+2)\Gamma (b +1)}
 \end{equation}
 which is larger than $2^a$ for any finite $b$ and $0< a <2$.
 The $2^a$ enhancement appears as the $n \to
  \infty$ limit of Eq.(\ref{rhon}).
 For small integer $n$, Eq.(\ref{rhon})  reduces
 to simple formulas obtained in refs. \cite{ddee,sssdd}.
 For $n=1$, we have 
 \begin{equation}
 \frac{{\cal F}_{\zeta}^{(b=1)}   (\zeta ) }{f(\zeta )}
  = \frac1{(1-a/2)(1-a/3)}  \, ,
  \label{rho1} 
 \end{equation}
 which gives the factor of 3 for the  enhancement if $a=1$.
 For $b=2$, the ratio (\ref{rhon}) becomes
 \begin{equation}
 \frac{{\cal F}_{\zeta}^{(b=2)}   (\zeta ) }{f(\zeta )}
  = \frac1{(1-a/3)(1-a/4)(1-a/5)}  \, ,
  \label{rho2} 
 \end{equation}
  producing a smaller enhancement factor $5/2$ for $a=1$. 
 Calculating the enhancement factors,
one should remember that the gluon SPD ${\cal F}_{\zeta}(X)$ 
reduces to $Xf_g(X)$ in the $\zeta =0$ limit. 
Hence, to get the enhancement factor corresponding to  the 
$ f_g(x) \sim x^{-\lambda}$  small-$x$ behavior of the forward 
gluon density, one should take $a= \lambda -1$ in Eq.(\ref{rhon}),
i.e., despite the fact that the $1/x$ behavior of the 
singlet quark distribution gives the factor of 3
for the ${\cal R}^{(1)} (\zeta)$ ratio, the same shape of 
the gluon distribution 
results in no enhancement.

 Due to evolution, the effective parameter $a$ 
characterizing the small-$x$ behavior 
of the forward distribution 
is an increasing function of $Q^2$.
As a result, for fixed $b$, the ${{\cal R}^{(b)} 
  (\zeta ) }$  
ratio  increases with $Q^2$.  
In general, the  profile of $\tilde f (\tilde x,  \alpha  )$
 in the $\alpha$-direction
is also  affected by the pQCD  evolution. 
In particular, in ref. \cite{ddee} it was shown that
  if one takes an ansatz  corresponding 
  to an extremely  asymmetric profile function $\rho (\alpha) 
  \sim  \delta (1+\alpha)$,  the shift 
  of the profile function to a more symmetric 
  shape  is clearly visible in the evolution 
  of the relevant SPD. 
In the next section, we  will study
 the interplay between 
 evolution of $x$ and  $\alpha$  profiles 
 of DDs.

\section{QCD evolution and profile of DDs}

Both the shape of the forward distributions
$\tilde f(x ; \mu)$ reflected in the $x$-dependence of the
DDs $\tilde f(x,\alpha ; \mu)$  and their
profile in the $\alpha$-direction 
are affected by the pQCD evolution.  At the one-loop
 level,  the solution for  QCD evolution
equations  is known in the operator form 
\cite{bbal},  so that choosing specific
matrix elements one can convert the universal 
solution into four (at least)    evolution patterns: 
for usual parton densities ($\langle p | \ldots |p \rangle$
case), distribution amplitudes $\langle 0 | \ldots |p \rangle$
case),   skewed and double  parton distributions 
($\langle p-r | \ldots |p \rangle$ case). 
Since all the types of the pQCD evolution 
originate from  the same source, one may  expect  an 
 interplay between  the $x$- and $\alpha$- 
  aspects of the DDs evolution.

In  the  simplest 
case of 
 flavor-nonsinglet (valence) functions, the
 multiplicatively renormalizable operators 
 were originally found in Ref.  \cite{tmf}
\begin{equation}
{\cal O}_n^{NS}   = (z\partial_+)^n \, \bar \psi 
\lambda^a \hat z C_n^{3/2} 
( z\stackrel{\leftrightarrow}{D}/z\partial_+) \psi \, .
 \label{100} \end{equation}
In contrast, the usual  operators $\bar \psi 
\lambda^a \hat z  
( z\stackrel{\leftrightarrow}{D})^n  \psi  $
mix under renormalization with the lower spin operators 
$ (z\partial_+)^{n-k} \bar \psi 
\lambda^a \hat z  
( z\stackrel{\leftrightarrow}{D})^k  \psi  $.
The symbolic  notation 
$ (z\stackrel{\leftrightarrow}{D}/z\partial_+)$  
with 
$\stackrel{\leftrightarrow}{D}
\equiv \stackrel{\rightarrow}{D} -
\stackrel{\leftarrow}{D}$ , $\partial_+ \equiv 
 \stackrel{\rightarrow}{D} +
\stackrel{\leftarrow}{D}
= \stackrel{\rightarrow}{\partial} +
\stackrel{\leftarrow}{\partial}$ 
and  $C_n^{3/2}(\alpha)$ being   the Gegenbauer polynomials
is borrowed from Ref.  \cite{tmf}. 
In Ref.  \cite{tmf} it was also noted 
that these  operators coincide with the  
free-field conformal tensors. 
As pointed  out in Ref. \cite{npd}, the multiplicative 
renormalizability
of ${\cal O}_n^{NS}$ operators means that 
 the   Gegenbauer moments 
\begin{equation}
{\cal C}_{n}^{NS}(\xi , \mu) = \xi^n \int_{-1}^1 C_n^{3/2} (z / \xi) \, 
H^{NS}( z, \xi; \mu) \, d z
\label{101} \end{equation}
of  the skewed parton distribution  
$H^{NS}(z, \xi;\mu)$ have a simple evolution \cite{npd}:
\begin{equation}
{\cal C}_{n}^{NS}(\xi, \mu)  = {\cal C}_{n}^{NS}(\xi, \mu_0)
\left [ \frac{ \ln \mu_0/\Lambda}{ \ln \mu/\Lambda} 
\right ]^{\gamma_n/ \beta_0} \,  ,
\label{102} \end{equation}
where  $\beta_0 = 11 -\frac23N_f$ is the 
lowest coefficient of the QCD $\beta$-function
and $\gamma_n$'s  are the nonsinglet anomalous dimensions \cite{gw,gp}.
Going from SPDs to DDs, writing the SPD variable  $\tilde x$ 
in terms of DD variables $ \tilde x = x + \alpha \xi $ and using 
\begin{equation}
C_n^{3/2} (x/\xi + \alpha) = \sum_{l=0}^{n} 
\frac{\Gamma(n-l+3/2)}{ \Gamma (3/2) (n-l)!}
 \, (2 x/\xi)^{n-l} \, 
C_l^{3/2+n-l} (\alpha) \,  ,
\end{equation}
one can express the Gegenbauer moments ${\cal C}_{n}(\xi, \mu)$ 
in terms of the combined 
[$x$-ordinary $\otimes $ $\alpha$-Gegenbauer]  moments 
of the relevant DDs:
\begin{equation}
{\cal C}_{n}^{NS}(\xi, \mu) 
= \sum_{k=0}^{[n/2]}  \xi^{2k} \,  
\int_{-1}^1 d x  \int_{-1+|x|}^{1-|x|}
2^{n-2k} \, \frac{\Gamma(n-2k+3/2)}{ \Gamma (3/2) (n-2k)!}
 \, x^{n-2k}
C_{2k}^{3/2+n-2k} (\alpha)\,  \, \tilde f^{NS} (x, \alpha;\mu) 
\, d \alpha \, . 
\label{gedd}
\end{equation}
Hence,  each $x^m C_l^{3/2+m}(\alpha)$ 
moment of $ \tilde f^{NS} (x, \alpha;\mu)$ is multiplicatively
renormalizable and its evolution
is governed by the anomalous dimension
$\gamma_{l+m}$ \cite{npd,ddee}.
In  Eq. (\ref{gedd}), we took into account that
 $\alpha$-DDs $\tilde f  (x,
\alpha)$ are  always even in $\alpha$, which  gives an expansion
of the Gegenbauer moments in powers of $\xi^2$. 
In the nonsinglet case, the Gegenbauer
moments ${\cal C}_{n}(\xi, \mu)$ are nonzero  
for  even $n$  only. A similar represenation can be written
for the Gegenbauer
moments of the 
singlet quark 
distributions. In the latter case,  the DD $\tilde f^{S}  (x, \alpha)$ 
is  odd in $x$, and only odd Gegenbauer
moments ${\cal C}_{n}^{S} (\xi, \mu)$ do not vanish.

Another simple case is the evolution 
of the gluon distributions in
pure gluodynamics. 
Then the  multiplicatively renormalizable operators with the same 
Lorentz spin 
$n+1$ as in Eq. (\ref{100}) 
 are 
\begin{equation}
{\cal O}^G_n = z^{\mu} z^{\nu} (z\partial_+)^{n-1} G_{\mu
\alpha} C_{n-1}^{5/2} 
( z\stackrel{\leftrightarrow}{D}/z\partial_+) \, G_{\alpha \nu} \, . 
\end{equation}
Due to the symmetry properties of gluon DDs, only  Gegenbauer moments 
\begin{equation}
{\cal C}_{n}^{G}(\xi , \mu) = \xi^{n-1} \int_{-1}^1 C_{n-1}^{5/2} (z / \xi) \, 
 H^{G}( z, \xi; \mu) \, d z
\label{1010} \end{equation}
with odd $n$ do not vanish.
 The  Gegenbauer moment  can also be 
written in terms of DDs:
\begin{equation}
{\cal C}_{n}^G(\xi, \mu) 
= \sum_{k=0}^{[(n-1)/2]}  \xi^{2k} \,  
\int_{-1}^1 d x  \int_{-1+|x|}^{1-|x|}
2^{n-2k-1} \, \frac{\Gamma(n-2k+3/2)}{ \Gamma (5/2) (n-2k-1)!}
 \, x^{n-2k}
C_{2k}^{3/2+n-2k} (\alpha)\,  \, \tilde f^G (x, \alpha) 
\, d \alpha \, . 
\label{geddglu}
\end{equation}
 Two shifts: $n \to n-1$ and $3/2 \to 5/2$  in some sense 
 compensate each other. Again, each   
 combined $x^m C_l^{3/2+m}(\alpha)$ moment 
 of $ \tilde f^G (x, \alpha)$ is multiplicatively
renormalizable and its evolution
is governed by the anomalous dimension
$\gamma_{l+m}^{GG}$ \cite{npd,ddee}.

Since the Gegenbauer polynomials 
$C_l^{3/2+m}(\alpha)$ are orthogonal with the 
weight  $(1- \alpha^2)^{m+1}$, evolution of 
the $x^m$-moments  of DDs 
in both cases is given by the  formula \cite{ddee} 
\begin{equation} 
\tilde f_m (\alpha\,  ;     \,  \mu)
\equiv \int_{-1}^1 x^m \tilde f (x, \alpha\,  ;     \,  \mu) \, dx = 
(1- \alpha^2)^{m+1} \sum_{k=0}^{\infty} A_{ml}
C^{m+3/2}_l(\alpha ) \left [\log (\mu /\Lambda) \right]^
{- \gamma_{m+l}/\beta_0} \,  , 
\label{eq:fnnons}
 \end{equation}
where the coefficients $A_{ml}$  
are proportional to  $x^m C_l^{3/2+m}(\alpha)$
moments of DDs.
A similar representation 
holds in the singlet case, with 
 $\left [\log (\mu /\Lambda) \right]^
{- \gamma_{m+l}/\beta_0}$ substituted by
a linear combination of  terms involving 
 $\left [\log (\mu /\Lambda) \right]^
{- \gamma_{m+l}^+/\beta_0}$ and  
$\left [\log (\mu /\Lambda) \right]^
{- \gamma_{m+l}^-/\beta_0}$ with  singlet 
anomalous dimensions $\gamma_{m+l}^{\pm}$ 
obtained by diagonalizing the coupled 
quark-gluon evolution equations \cite{ddee}.

The anomalous dimensions $\gamma_{n}$ increase with raising $n$,
and, hence,   the $m$th $x$-moment of 
$ \tilde f (x, \alpha;\mu )$ is  asymptotically 
dominated by the $\alpha$-profile 
$(1-\alpha^2)^{m+1}$. 
Such a correlation between $x$-  and 
$\alpha$-dependences
of $ \tilde f (x, \alpha;\mu )$  is 
not something exotic. 
 Take a DD which is  constant in its  support region.
 Then  its   $x^m$-moment behaves like 
 $(1- |\alpha|)^{m+1}$, i.e., the  width of the 
 $\alpha$ profile   decreases  with 
 increasing $n$.
 This result is easy to understand:
 due to the spectral condition $ |\alpha| \leq 1 -|x|$,
 the $x^m$ moments with larger $m$ are 
 dominated by regions which are narrower
 in the $\alpha$-direction.

 These   observations
suggests to try  a model in which the moments 
 $ \tilde f_m   (\alpha; \mu)$ has the asymptotic $(1-\alpha^2)^{m+1}$
 profile
even at non-asymptotic  $\mu$. This is equivalent
to assuming that all the combined moments
$x^m C_l^{3/2+m}(\alpha)$ with $l>0$ vanish. 
Note that this assumption 
is stable wrt pQCD evolution. 
Since integrating $\tilde f_m (\alpha\,  ;     \,  \mu) $
over $\alpha$  one should get the moments $ \tilde f_m (\mu) $ of the forward 
density $f(x ; \mu)$, the DD moments  $ \tilde f_m   (\alpha; \mu )$ 
in this model are given by 
   \begin{equation}
  \tilde f_m   (\alpha; \mu )   
   = \rho_{m+1} (\alpha)
   \tilde f_m (\mu)  
   \end{equation}
 where  $\rho_{m+1} (\alpha) $ is  the normalized profile 
function   (cf. Eq.(\ref{rhon})). In explicit form: 
 \begin{equation}
  \int_{-1}^{1} x^m  
\tilde f(x,\alpha\,  ;     \,  \mu) \, dx \, 
 =
 \frac{\Gamma(m+5/2)}{\Gamma (1/2) \, (m+1)!}
  (1-\alpha^2)^{m+1}  
  \int_{-1}^{1} \tilde f (z;\mu) z^m dz \  .
\label{10810} 
\end{equation}
 In this relation,  all the dependence on  $\alpha$ 
 can be trivially shifted 
to the  lhs of this equation, and we immediately see
that  
$\tilde f(x,\alpha\,  ;     \,  \mu)$  in this model 
is a function of $x/(1-\alpha^2)$:
  \begin{equation}
 \tilde f(x,\alpha\,  ;     \,  \mu)  = F(x/(1-\alpha^2);\, \mu ) \, 
 \theta (0 < x/(1-\alpha^2 ) < 1) \, .
  \end{equation} 
A direct   relation  
between $\tilde f(z,\mu)$ and $F(u;\mu)$ 
can be easily obtained using the basic fact that
integrating $\tilde f(x,\alpha\,  ;     \,  \mu)$ 
 over $\alpha$ one should get  
the forward density $\tilde f(z,\mu)$;
 e.g., for positive $z$ we have 
 \begin{equation}
 f(z) = z  \int_z^1 \frac{F(u)}{u^{3/2} \sqrt{u-z}} \, du  \, . 
 \end{equation} 
This relation  has the structure of the Abel equation.
Solving it for $F(u)$ we get 
\begin{equation}
F(u) = - \frac{u^{3/2}}{\pi} \int_u^1 
\frac{ \left [ f(z)/z \right ]^{\prime}}{\sqrt{z-u}}  \, dz \,  .
\label{inabel}
\end{equation}
Thus, in this model, knowing the forward  density $f(z)$ 
one can calculate the double distribution function 
$\tilde f(x,\alpha)  = F (x/(1-\alpha^2))$.

Note, however, that the model derived above  
violates  the DD support condition $|x|+|\alpha| \leq 1$:
the restriction $|x| \leq 1- \alpha^2$  
defines a larger area. Hence, the model
is only applicable in a situation
when the difference between two 
spectral conditions can be neglected. 
A practically important case is the shape of 
$ H(\tilde x, \xi )$  for small $\xi$. 
Indeed, calculating  $H(\tilde x, \xi ) $
 for small $\xi$ one integrates  the relevant
 DDs $\tilde f(\tilde x)$ over practically 
 vertical lines. 
 If $\tilde x$ is also small,  both the correct 
$|\alpha| \leq 1 - |x|$ and model $\alpha^2 \leq 1- |x|$ 
conditions 
can be substituted by $|\alpha| \leq 1$. 
Now,  if $ \tilde x \gg \xi$, a   slight deviation
of the integration line from  the vertical 
direction can be neglected and $H(\tilde x, \xi ) $
can be approximated by the forward limit $\tilde f(\tilde x)$.

 Specifying  the  ansatz for $f(z)$, one 
  can  get an explicit expression for the model DD by 
 calculating $F(u)$ from Eq. (\ref{inabel}).  
However, in the  simplest  case when $f(x) =  A x^{-a}$ 
for small 
$x$, the result is evident without any  calculation:
the DD $f(x,\alpha)$ which is a function of the
ratio $x/(1-\alpha^2)$ and reduces to $A x^{-a}$
after in integration over $\alpha$ must be given by 
$$f(x,\alpha)  = \rho_{a} (\alpha) f(x)$$   
where $\rho_{a} (\alpha)$ is  the normalized profile
function of Eq.(\ref{rhon}):  
\begin{equation}
f(x,\alpha)  =   
A \, \frac{\Gamma(a+5/2)}{\Gamma (1/2) \, \Gamma (a+2)}
  (1-\alpha^2)^{a}  \, x^{-a}  .
  \end{equation} 
 This DD   is 
 a particular  case of the general
 factorized ansatz  $f(x,\alpha)  = \rho_{n} (\alpha) f(x)$
 considered in the previous section. 
 Its most nontrivial feature is the correlation
 $n=a$ between the   profile function parameter $n$ 
 and the power $a$ characterizing the small-$x$
 behavior of the forward distribution.  
 
    Knowing DDs, the relevant SPDs $H(\tilde x, \xi )$  
can be obtained in the standard way from   $\tilde f(x,\alpha)$
for  quarks and from  $x \tilde f^G (x,\alpha)$ 
in the case of   gluons. 
In particular,  the SPD enhancement  factor ${\cal R} (\zeta)$ 
for small $\zeta$ 
in this  model is given by 
\begin{equation}
 \frac{{\cal F}_{\zeta}^{Q}   (\zeta ) }{f^Q(\zeta )}
  = \frac{\Gamma (2a+2)}  
  {\Gamma (a+2)\Gamma (a +1)}
 \end{equation}
for quarks and  by 
\begin{equation}
 \frac{{\cal F}_{\zeta}^{G}   (\zeta ) }{\zeta f^G (\zeta )}
  = \frac{\Gamma (2a+2)}  
  {\Gamma (a+3)\Gamma (a +1)}
 \end{equation}
for gluons.

The use of the asymptotic profiles 
for DD  moments $\tilde f_n (\alpha)$ 
is the basic assumption of the  model 
described above.  However, if one is interested 
in SPDs  
for small $\xi$, the impact of 
deviations  of  $\tilde f_n (\alpha)$ from 
the asymptotic profile is suppressed. 
Even if the higher harmonics are  present 
in $\tilde f_n (\alpha)$, i.e.,  if  
the $x^{n-2k} C_{2k}^{3/2+n-2k} (\alpha)$
 moments  of $\tilde f(x,\alpha)$ 
 are nonzero for $k \geq 1$ values, their contribution
 into the Gegenbauer moments  ${\cal C}_n (\xi, \mu) $ 
  is strongly suppressed by $\xi ^{2k}$ factors [see Eq.(\ref{gedd})].
  Hence, for small $\xi$, 
  the shape of $H(\tilde x, \xi )$  
  for a wide variety of model $\alpha$-profiles 
   is very close to that based  on the asymptotic profile model. 
   
   Absence of   higher harmonics in $\tilde f_n (\alpha)$
   is equivalent to absence of the $\xi$-dependence 
   in the Gegenbauer moments ${\cal C}_n (\xi, \mu) $.
   The assumption that   ${\cal C}_n (\xi, \mu) $  
   do not depend on $\xi$ was the starting point 
   for the model of SPDs $H(\tilde x, \xi )$ 
   constructed in ref. \cite{sgmr}. 
   Though the formalism of DDs was not 
   used in ref. \cite{sgmr},  
   both approaches lead to identical  results:
   the final result of \cite{sgmr} has the form 
   of a DD representation for $H(\tilde x, \xi )$.
   In  Appendix A, we also start with 
    ${\cal C}_n (\xi, \mu) = 
   {\cal C}_n (0, \mu)$  and rederive   the 
   DD corresponding to the asymptotic profile model.

\section{Evolution Algorithm} 

At one loop,  evolution equations for nonforward parton distributions
${\cal F}_\zeta^a (x,\mu)$ can be written as  
\begin{eqnarray}
 \mu {d {\cal F}_\zeta^a (x,\mu) \over d \mu} &=&
 \sum_b \int\limits_0^1  W_\zeta^{ab}(x,z) {\cal F}_\zeta^b (z,\mu^2)
 \, dz 
\label{EvEqn}
\end{eqnarray}
or in a ``matrix''  notation 
\begin{eqnarray}
\mu {d \hat{\cal F}_\zeta^a (\mu) \over d \mu} &=&
  \hat W_\zeta^{ab}
 \otimes  \hat{\cal F}_\zeta^b(\mu) \, .
\label{formalEvEqn}
\end{eqnarray}

Using  the 
explicit one-loop form of the effective coupling constant
\begin{equation} 
\alpha_s (\mu) = \frac{4 \pi}{\beta_0 \log (\mu^2 / \Lambda^2) }
\end{equation}
($\beta_0 = 11- \frac23 N_f$) and the symbolic 
notations of  (\ref{formalEvEqn}), one can present the formal solution
for the set of evolution equations in the form of an expansion
\begin{eqnarray}
\hat{\cal F}_\zeta(Q^2) &=& 
\exp \left (L(Q^2,{Q_0}^2)\ \hat W_\zeta \right )
                    \otimes 
       \hat{\cal F}_\zeta({Q_0}^2) \\
 &=& \sum_n {\left (L(Q^2,{Q_0}^2\right )^n\over n!} 
\left [ \left (\hat W_\zeta  \right )^n
\otimes \hat{\cal F}_\zeta({Q_0}^2) \right ] \, . 
\label{FormalSol}
\end{eqnarray}
To get the  $n=1$ term 
$\delta_1 \hat{\cal F}_\zeta \equiv
\hat W_\zeta \otimes \hat{\cal F}_\zeta({Q_0}^2)$ of this expansion,
we evaluate numerically the convolution of the 
kernel $\hat W_\zeta$ with the initial   distributions 
$\hat{\cal F}_\zeta({Q_0}^2)$.    
To get the $n=2$ term $\delta_2 \hat{\cal F}_\zeta \equiv
(\hat W_\zeta )^2  \otimes \hat{\cal F}_\zeta({Q_0}^2)$,
 we convolute $\hat W_\zeta$ 
with the smoothly  interpolated
result  of the first iteration $\delta_1 \hat{\cal F}_\zeta$,
and so on.  
 After obtaining $\delta_1 \hat{\cal F}_\zeta$, 
 $\delta_2 \hat{\cal F}_\zeta$, etc., we  construct 
 evolved 
    distributions $\hat{\cal F}_\zeta({Q}^2)$ 
for any desired value of $Q^2$. 
Of course, the number of necessary  
iterations of $\hat W_\zeta$ with 
the  initial distributions $\hat{\cal F}_\zeta({Q_0}^2)$
depends on the 
size of  the expansion parameter 
\begin{eqnarray}
L(Q^2,{Q_0}^2) &=& {2 \over \beta_0}
\log \left [ {\log \left (Q^2/\Lambda^2 \right )\over
             \log \left ({Q_0}^2/\Lambda^2\right )}\right ] \, .
\label{LqqLgg}
\end{eqnarray}
When $L$ is not very large, it is sufficient to
calculate  just one or two 
iterations.

\section{Evolution of nonforward quark distributions}

The evolution of skewed parton distributions 
was studied numerically in refs. \cite{ffgs,bgms,lech,mr,gmr}. 
In this section, we perform the numerical evolution
of SPDs using the algorithm described in the previous session
and present the
results  illustrating  evolution 
patterns for SPDs constructed 
using  the factorized model (\ref{65n}) with
different choices for the profile function 
$h(x,\alpha)$.   

As we discussed earlier,
the use of the infinitely narrow profile
function  $h(x,\alpha) = \delta (\alpha)$ 
gives the simplest model in which  
OFPDs $H(\tilde x,\xi)$ coincide 
with forward distributions $\tilde f(\tilde x)$. 
In terms of NFPDs, this modell looks less trivial.
It gives  $
F^{(\infty)} (x,y) =  \delta(y - \bar x /2) f(x) $
for the $y$-DDs  which results in 
(untilded)  NFPDs  given by  shifted 
forward distributions
$${\cal F}_{\zeta} (X) =
{1\over(1-\zeta/2)} f\left({X-\zeta/2\over 1-\zeta/2}\right). $$
For any monotonic function $f(x)$ this gives 
NFPDs ${\cal F}_{\zeta} (X)$ 
which are larger in the region $X \geq \zeta $ than their 
forward counterparts. Due to  pQCD evolution,
$f(x,Q^2)$ get steeper in small-$x$ region,
i.e. the NFPDs  become even more strongly enhanced. 

\begin{figure}[htb]
\mbox{
   \epsfxsize=8cm
 \epsfysize=5cm
 \hspace{0cm}  
  \epsffile{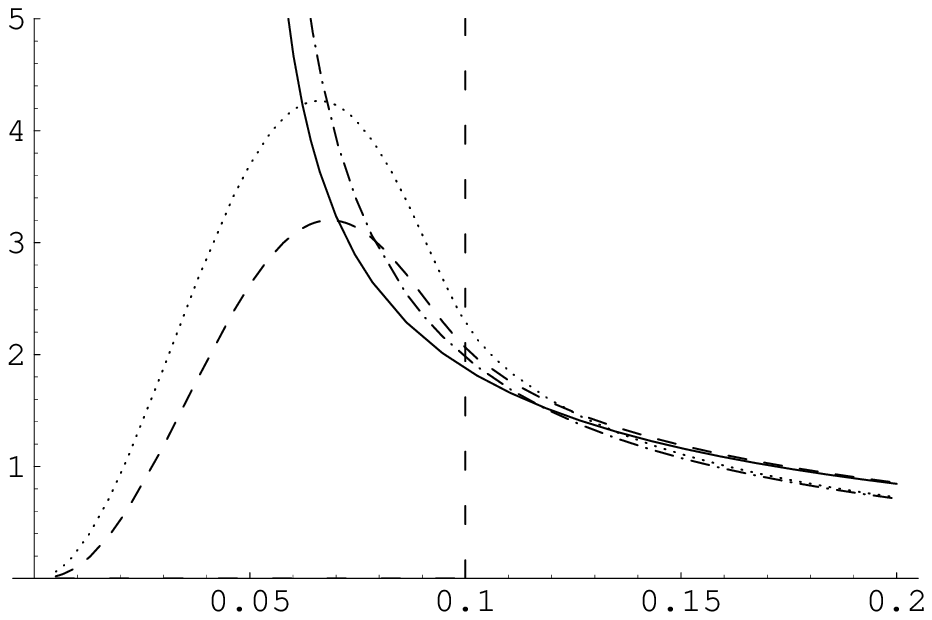}  } \hspace{0cm}
\mbox{
   \epsfxsize=8cm
 \epsfysize=5cm
 \hspace{0cm}  
  \epsffile{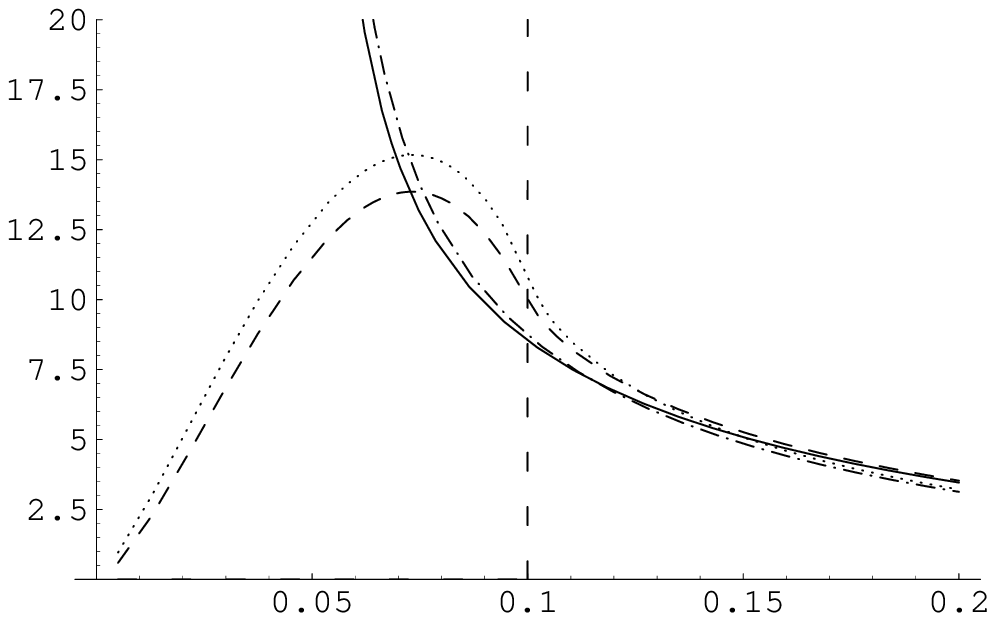}  }
{\caption{\label{fw-vs-nonfw-nons}
YM gluon (left) and quark nonsinglet (right) 
nonforward distributions $F^{G,NS}_\zeta(x)$
based on our model with $h_{as}$ profiles versus ``shifted'' forward ones
${1\over(1-\zeta/2)} f\left({x-\zeta/2\over 1-\zeta/2}\right)$,
at two different $Q^2$ scales: $1.5 \,{\rm GeV}^2$ (dashed and solid curves)
and $20{\rm GeV}^2$ (dotted and dash-dotted curves).  }}
\end{figure}

\begin{figure}[htb]
\mbox{
   \epsfxsize=8cm
 \epsfysize=5cm
 \hspace{0cm}  
  \epsffile{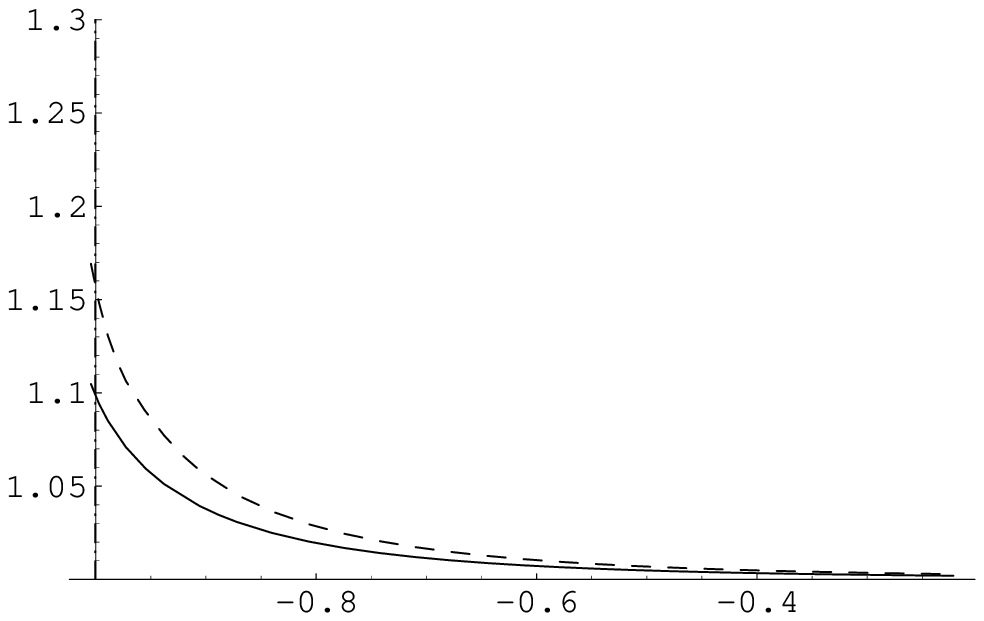}  } \hspace{0cm}
\mbox{
   \epsfxsize=8cm
 \epsfysize=5cm
 \hspace{0cm}  
  \epsffile{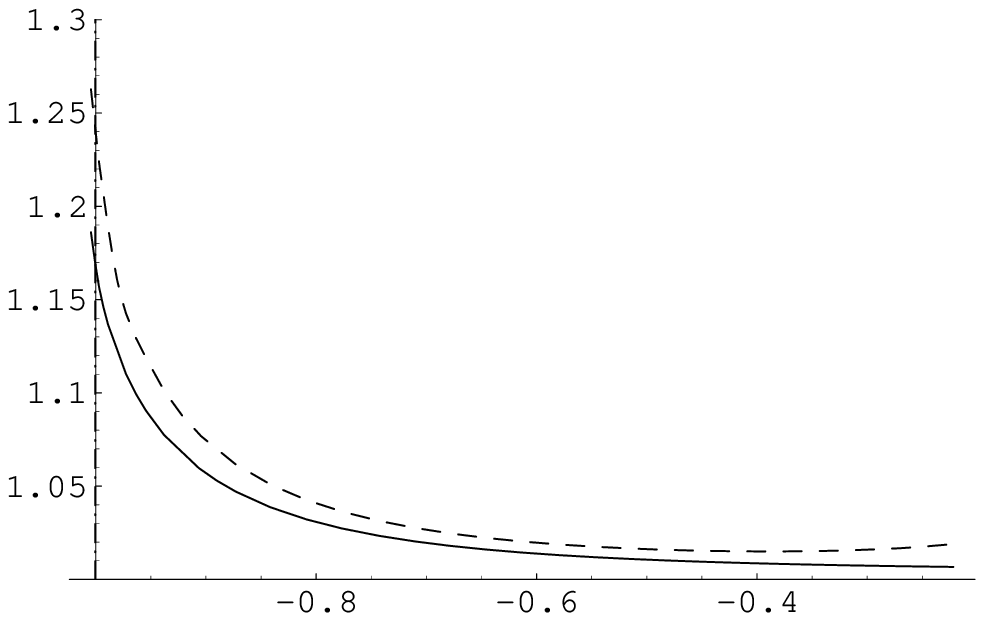}  }
{\caption{\label{fw-vs-nonfw-nons2}
Ratio of YM gluon (left) and nonsinglet quark (right) nonforward distributions
$F^{G,NS}_\zeta(x)$ obtained from our model with $h_{as}$ profiles to the
``shifted'' forward ones ${1\over(1-\zeta/2)} f\left({x-\zeta/2\over
1-\zeta/2}\right)$, at two different $Q^2$ scales: $1.5 \, {\rm GeV}^2$
(solid) and $20 \, {\rm}GeV^2$ (dashed);  $\zeta=0.1$.
   }}
\end{figure}

As noted in Section IV, the use of wide profile functions
 results in stronger enhancement for 
NFPDs in $X \geq \zeta $ region. 
For llustration see Figs.\ref{fw-vs-nonfw-nons},\ref{fw-vs-nonfw-nons2}. 
 In Fig.  \ref{fw-vs-nonfw-nons} we show $\zeta =0.1$ NFPDs 
 obtained ``by shift'' from forward distributions
 $f(x,Q^2)$  taken at two $Q^2$ values
 1.5 GeV$^2$ and 20 GeV$^2$. 
 For comparison, we show also NFPDs obtained 
 from $f(x, Q^2 =1.5\,{\rm GeV}^2)$ 
 using the $y$-DDs  with ``asymptotic'' profile functions
 $F_{as}^{NS}(x,y) = 6 y (1-x-y) f^{NS}(x) /(1-x)^3$ for nonsinglet 
 quark distributions and  $F_{as}^{YM}(x,y) = 30 y^2 (1-x-y)^2/(1-x)^5 f^G (x)$
 for gluon distibutions in pure gluodynamics.
Specifically, we took $f^{NS} (x, Q^2 =1.5\,{\rm GeV}^2) = \frac{35}{32}
x^{-1/2} (1-x)^3 $and $f^G (x, Q^2 =1.5\,{\rm GeV}^2) = 0.4
x^{-0.3} \ln (1/x) (1-x)^5 $. 
 The NFPDS constructed in this way were
 then numerically evolved to $Q^2 = 20\,{\rm GeV}^2$  
 using the approach outlined in Section VI
 and kernels given Appendix B.  
 The ratio of NFPDs obtained using these two models 
 is shown in Fig.\ref{fw-vs-nonfw-nons2}.
 As expected, the ratios increase with $Q^2$.

\begin{figure}[t]
\hspace{-0.5cm}
\mbox{
   \epsfxsize=8cm
 \epsfysize=5cm
 \hspace{0cm}  
  \epsffile{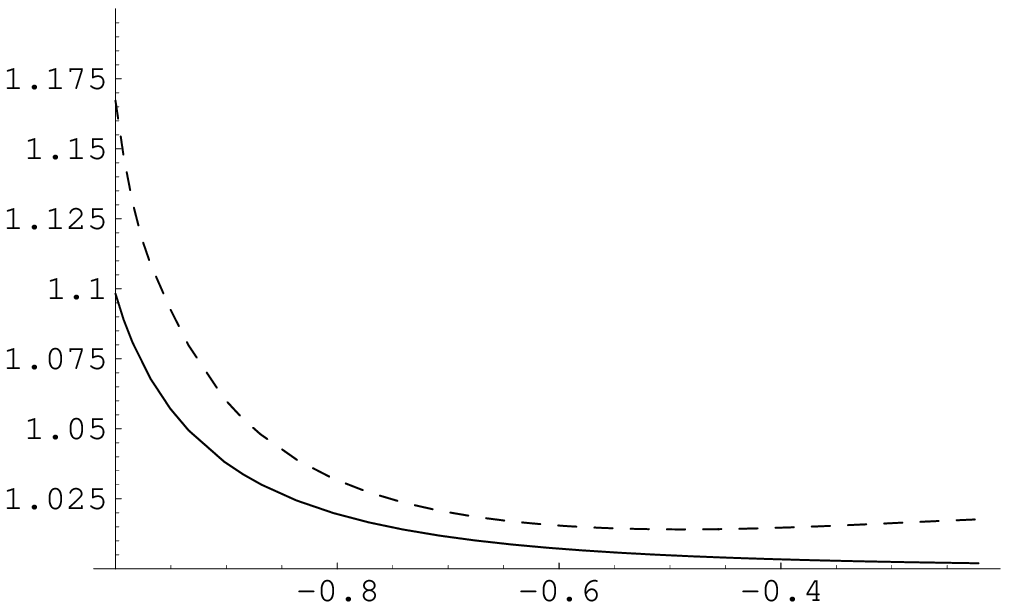}  } \hspace{0cm}
\mbox{
   \epsfxsize=8cm
 \epsfysize=5cm
 \hspace{0cm}  
  \epsffile{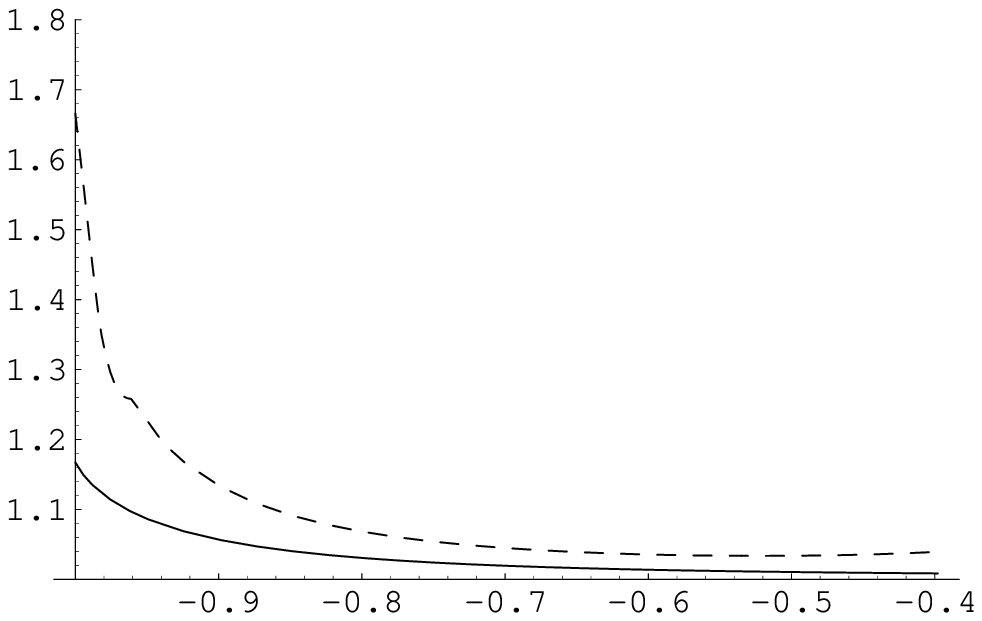}  } \\
\mbox{
   \epsfxsize=8cm
 \epsfysize=5cm
 \hspace{0cm}  
  \epsffile{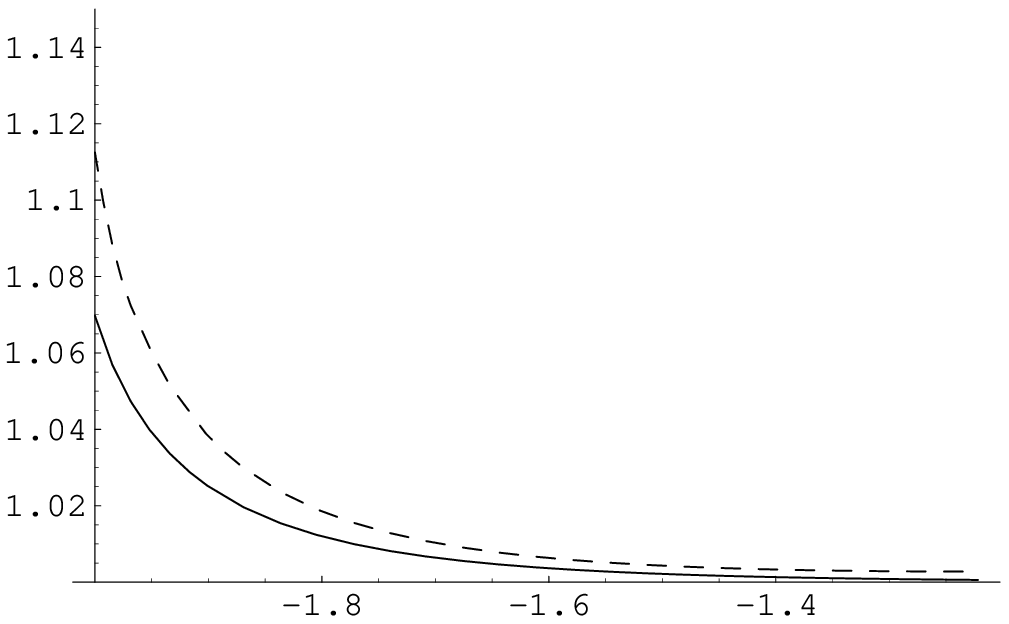}  } \hspace{0cm}
\mbox{
   \epsfxsize=8cm
 \epsfysize=5cm
 \hspace{0cm}  
  \epsffile{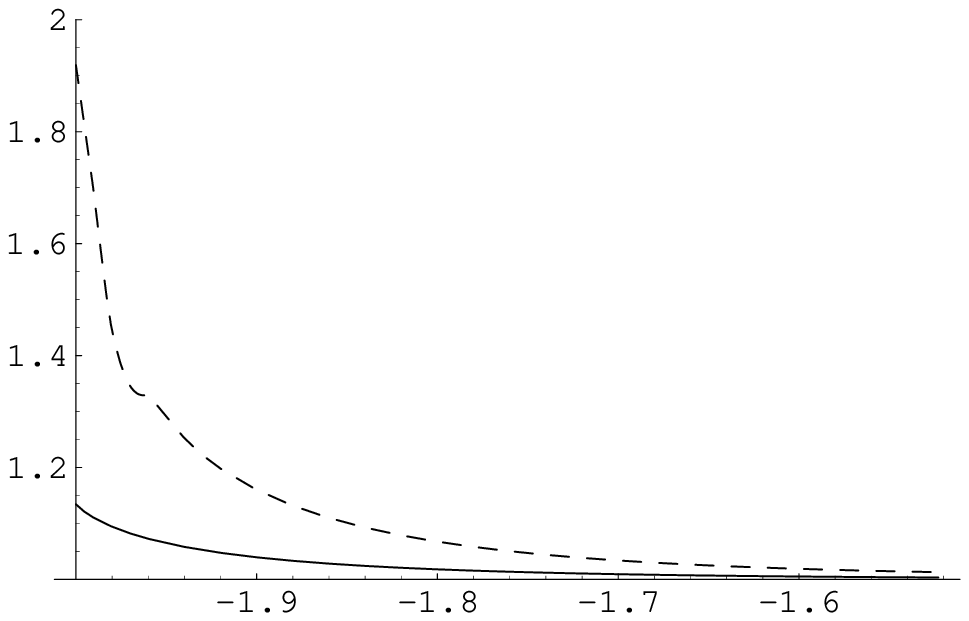}  } \\
\mbox{
   \epsfxsize=8cm
 \epsfysize=5cm
 \hspace{0cm}  
  \epsffile{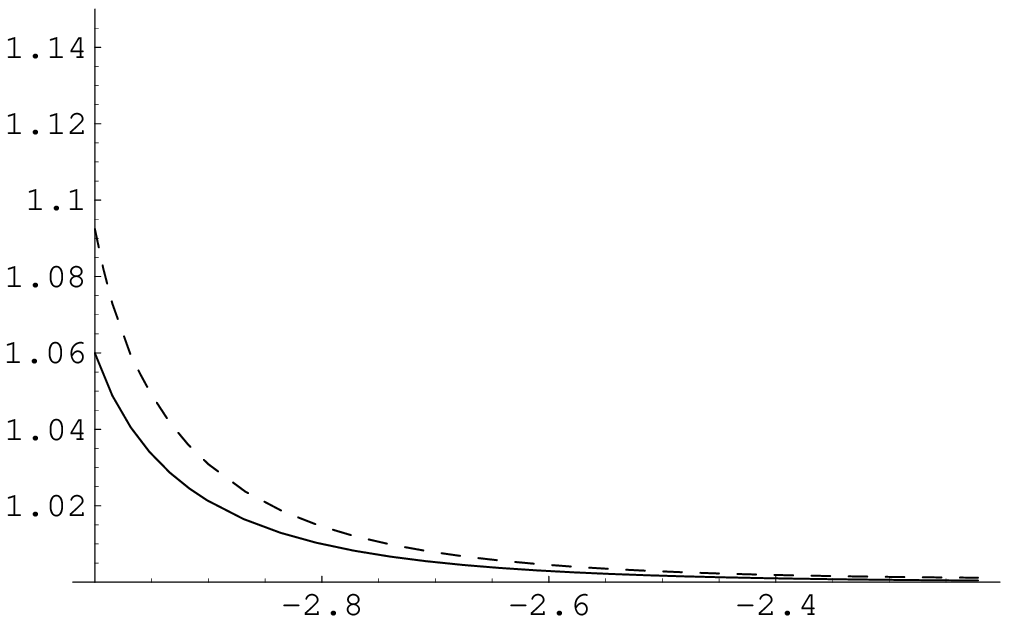}  } \hspace{0cm}
\mbox{
   \epsfxsize=8cm
 \epsfysize=5cm
 \hspace{0cm}  
  \epsffile{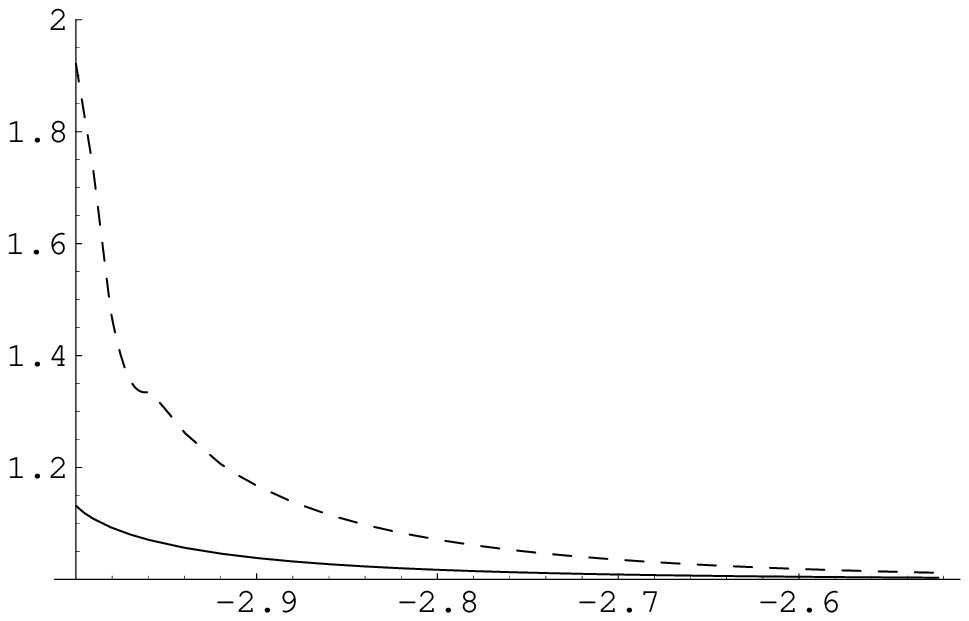}  } \\
{\caption{\label{fw-over-nonfw-s}
Ratio of QCD gluon (left) and singlet quark (right) nonforward distributions
$F^{G,S}_\zeta(x)$ obtained from our model with $h_{as}$ profiles to the
``shifted'' forward ones ${1\over(1-\zeta/2)} f\left({x-\zeta/2\over
1-\zeta/2}\right)$, at two different $Q^2$ scales: $1.5\,{\rm GeV}^2$ (solid)
and $20\,{\rm GeV}^2$ (dashed);  $\zeta=0.1$ - top, $\zeta=0.01$ - middle,
$\zeta=0.001$ - bottom.    }}
\end{figure}

The same construction can be performed 
in the singlet case. 
Main observation here is large enhancement 
for singlet quark distributions at $Q^2=20$\,GeV$^2$,
with the ${\cal F}_{\zeta}^S(\zeta)/f^S(\zeta/2)$ ratio
being close to 1.8, see Fig.\ref{fw-over-nonfw-s}.  
This is again in agreement with the estimates 
made in Section IV for $a \approx 1$.

\begin{figure}[htb]
\mbox{
   \epsfxsize=8cm
 \epsfysize=5cm
 \hspace{0cm}  
  \epsffile{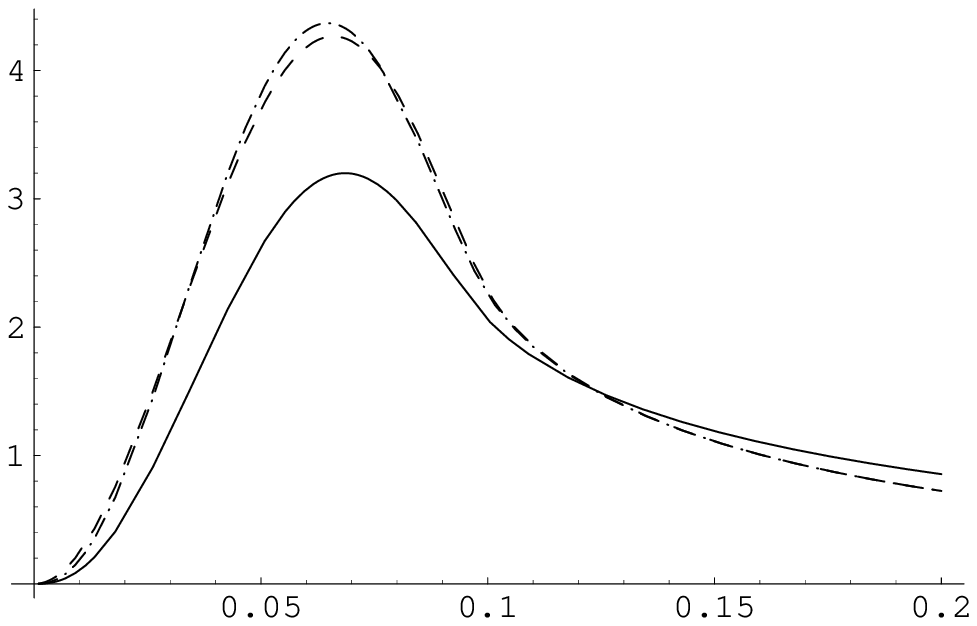}  } \hspace{0cm}
\mbox{
   \epsfxsize=8cm
 \epsfysize=5cm
 \hspace{0cm}  
  \epsffile{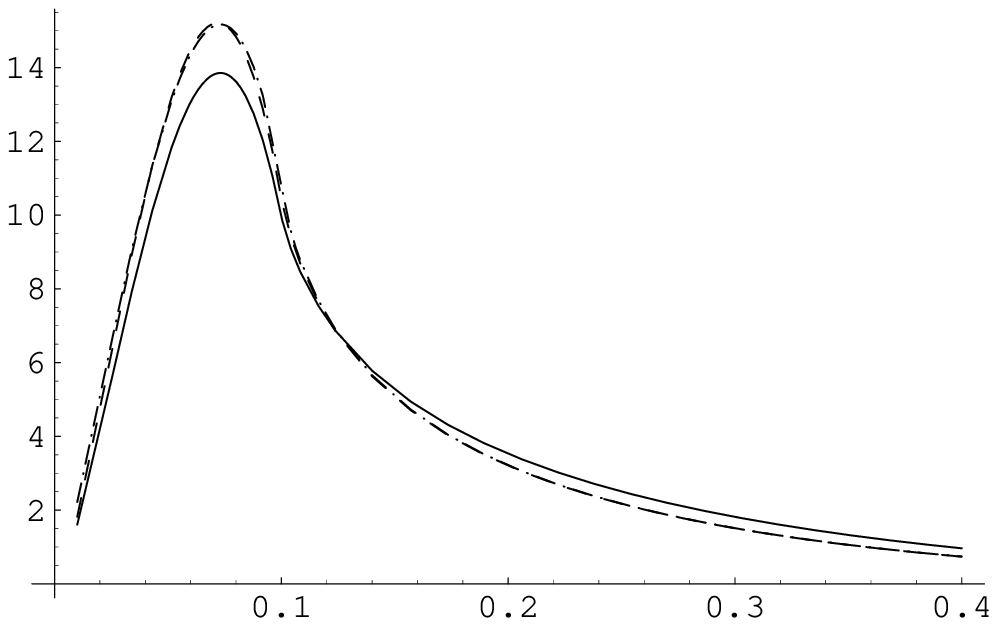}  }
{\caption{\label{Evolved-model}
Evolution of the YM gluon (left) and nonsinglet quark (right) 
nonforward distributions $F^{G,q}_\zeta(x)$
obtained in our model with $h_{as}$ profiles.
Solid curves correspond to the initial distributions at $Q^2=1.5\,{\rm GeV}^2$.
Dashed curves represent nonforward distributions evolved to $Q^2=20\,{\rm GeV}^2$,
dash-dotted ones are obtained from the model with the same  $h_{as}$ profiles
and with forward distributions evolved to $Q^2=20\,{\rm GeV}^2$.
   }}
\end{figure}

In the above examples
of nonsinglet quark distributions and gluon distributions
in pure gluodynamics  we took  ``asymptotic'' profiles.
It is interesting to test whether these profiles are
really stable under pQCD evolution. 
To this end, we compared two models for
$Q^2=20\,$GeV$^2$ distributions.
First, we took the forward distribution evolved to
$Q^2=20\, $GeV$^2$  and constructed model NFPD 
using the ``asymptotic'' profile. 
Second way is to construct NFPD from $Q^2=1.5\, $GeV$^2$ 
forward distribution using the asymptotic profile 
and then evolve NFPD to $Q^2=20\, $GeV$^2$ using nonforward kernels.
Fig. \ref{Evolved-model} shows that the results 
obtained in the two ways are practically identical.

\begin{figure}[t]
\begin{center}
\mbox{
   \epsfxsize=8cm
 \epsfysize=5cm
 \hspace{0cm}  
  \epsffile{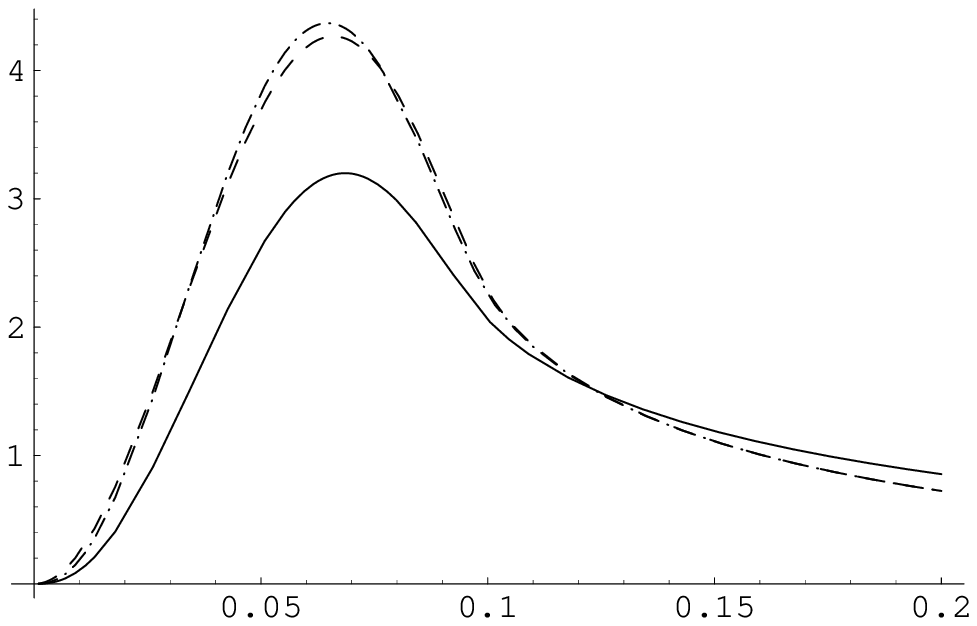}  } \\
\vspace{1cm}
\mbox{
   \epsfxsize=8cm
 \epsfysize=5cm
 \hspace{0cm}  
  \epsffile{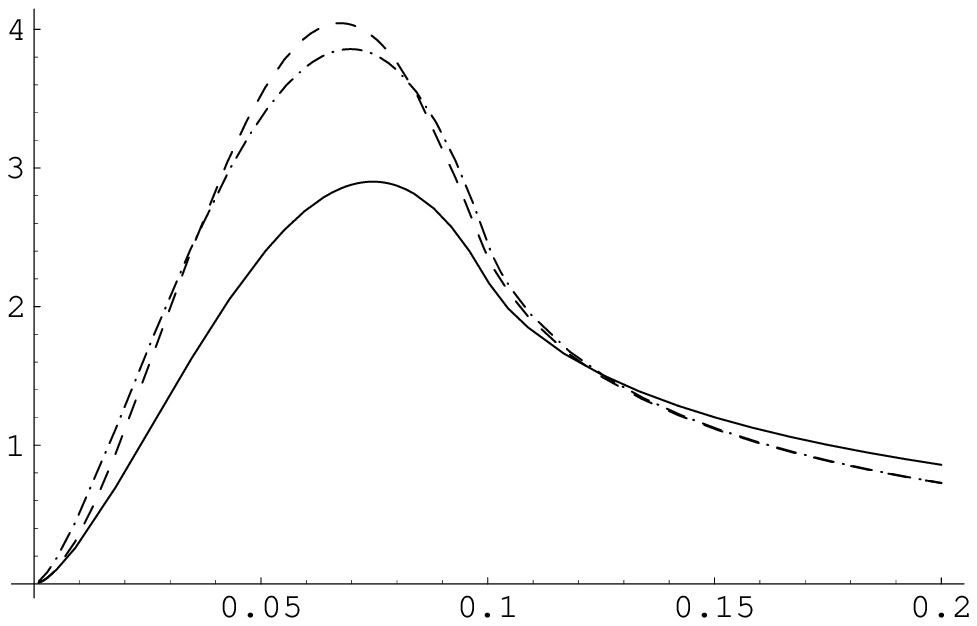}  } \\
\vspace{1cm}
\mbox{
   \epsfxsize=8cm
 \epsfysize=5cm
 \hspace{0cm}  
  \epsffile{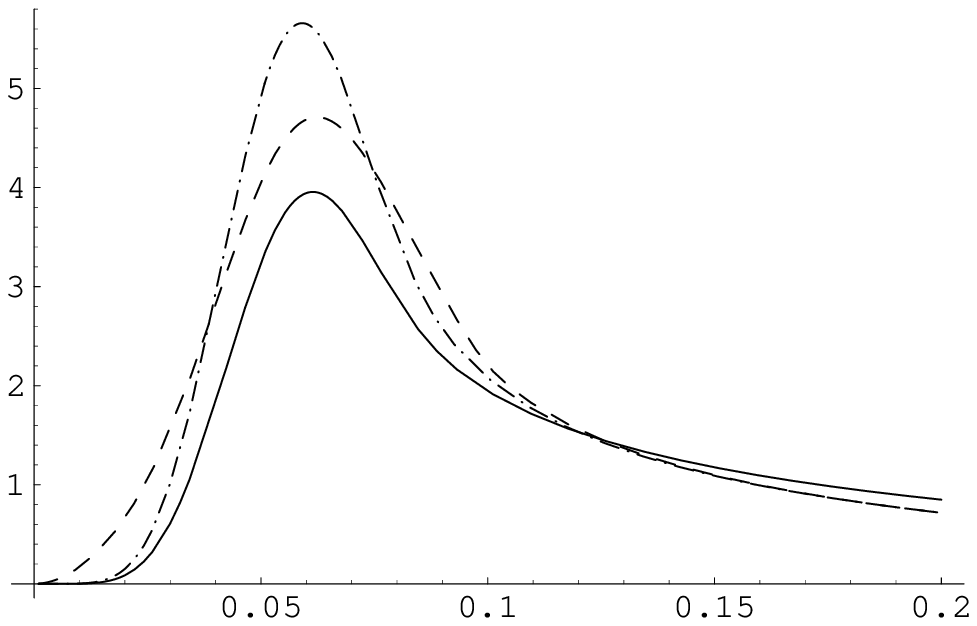}  }
\end{center}
{\caption{\label{Different-models-nons} 
The same as on previous set of figures (gluon distributions only) but
with different profiles $h(x,y)= N_\lambda(x) (y(1-x-y))^\lambda$: asymptotic
($\lambda=2$) - top, wide ($\lambda=1$) - middle, narrow  ($\lambda=6$) -
bottom.
   }}
\end{figure}

However, if one takes profiles strongly differing from the ``asymptotic''
ones, the curves obtained in the two ways described above,
visibly differ from each other, see Fig.\ref{Different-models-nons}.
 In the case of a wide profile  function,
the evolved NFPD looks like that constructed 
from evolved forward distribution but using a narrower profile.
 In other words,  the pQCD evolution in this case
 narrows the profile  function. Alternatively,
 if one starts with a too narrow 
 profile, then the evolved NFPD resembles  the model
 function  constructed 
from evolved forward distribution but using a wider profile.

\noindent

\begin{figure}[t]
\begin{center}
\mbox{
   \epsfxsize=8cm
 \epsfysize=5cm
 \hspace{0cm}  
  \epsffile{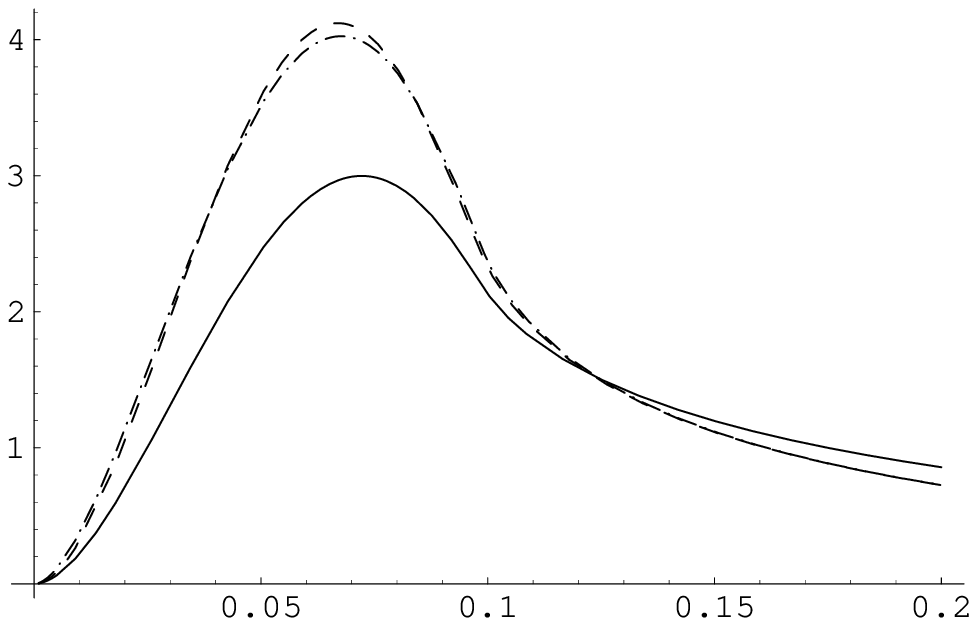}  } \\
\vspace{1cm}
\mbox{
   \epsfxsize=8cm
 \epsfysize=5cm
 \hspace{0cm}  
  \epsffile{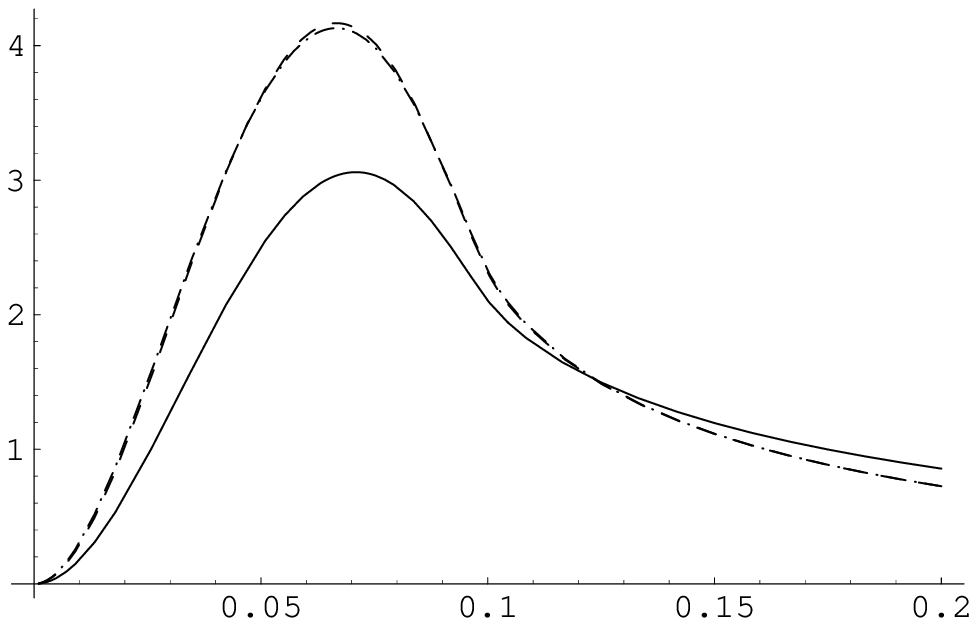}  } \\
\vspace{1cm}
\mbox{
   \epsfxsize=8cm
 \epsfysize=5cm
 \hspace{0cm}  
  \epsffile{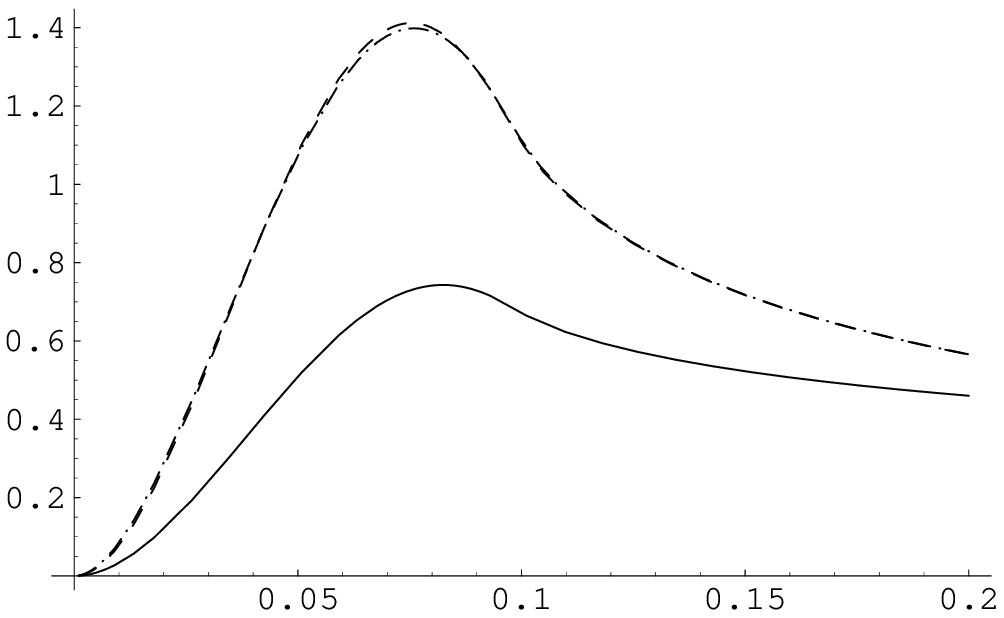}  }
\end{center}
{\caption{\label{Scientific-models-nons} 
Top and middle: the same as on previous set of figures but
with other profiles $h(x,y)= N_\lambda(x) (y(1-x-y))^\lambda$:
$\lambda=1.3$ - top, $\lambda=1.5$ - middle. Bottom figure
represents the evolution of the model obtained with the
profile parameter $\lambda=1.3$ and model forward distribution
$f_G(x) = 1/x^{0.3}$ (no logarithm).
   }}
\end{figure}

The study performed  in Section V (see also Appendix A)
demonstrated that at large $Q^2$ there should be 
a correlation between the $x$-dependence of 
the forward distributions and the form of the profile function.
Taking the  GRV-type parametrization  for 
gluon (with $f_g(x) \sim x^{-0.3} \ln (1/x) $) and quark singlet distributions 
at $Q^2=1.5$GeV$^2$,  we again compared  the $Q^2=20$GeV$^2$
curves constructed in two ways described above. 
A  better  agreement between the two 
models was observed for $\lambda =1.5$ rather than 
for $\lambda =1.3$.  However, the $\lambda =1.3$ profile
works perfectly if one takes the model with 
purely power-like behavior of the gluon distribution
$f_g(x) \sim x^{-0.3}$, see Fig.\ref{Scientific-models-nons}.

\noindent

\begin{figure}[htb]
\mbox{
   \epsfxsize=8cm
 \epsfysize=5cm
 \hspace{0cm}  
  \epsffile{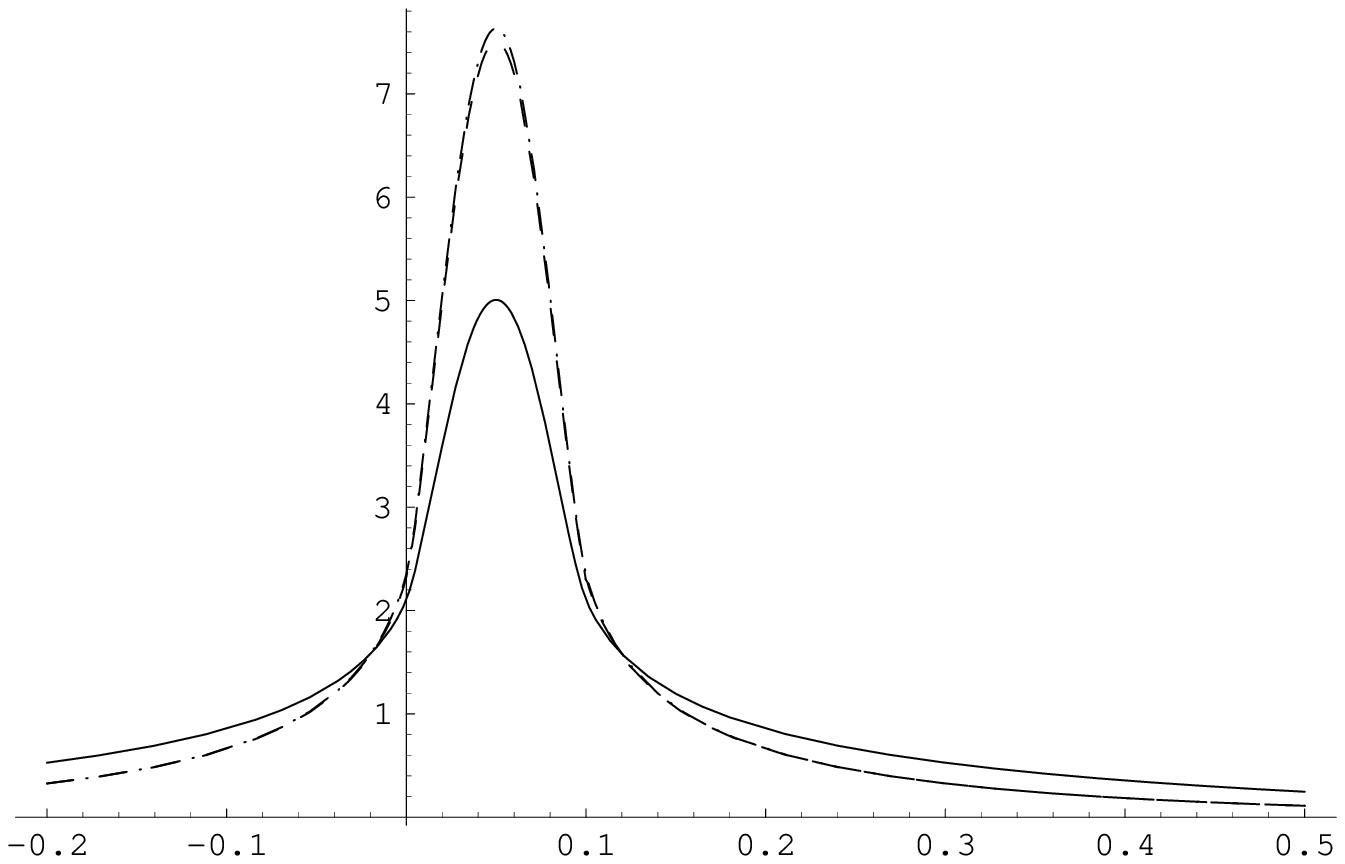}  } \hspace{0cm}
\mbox{
   \epsfxsize=8cm
 \epsfysize=5cm
 \hspace{0cm}  
  \epsffile{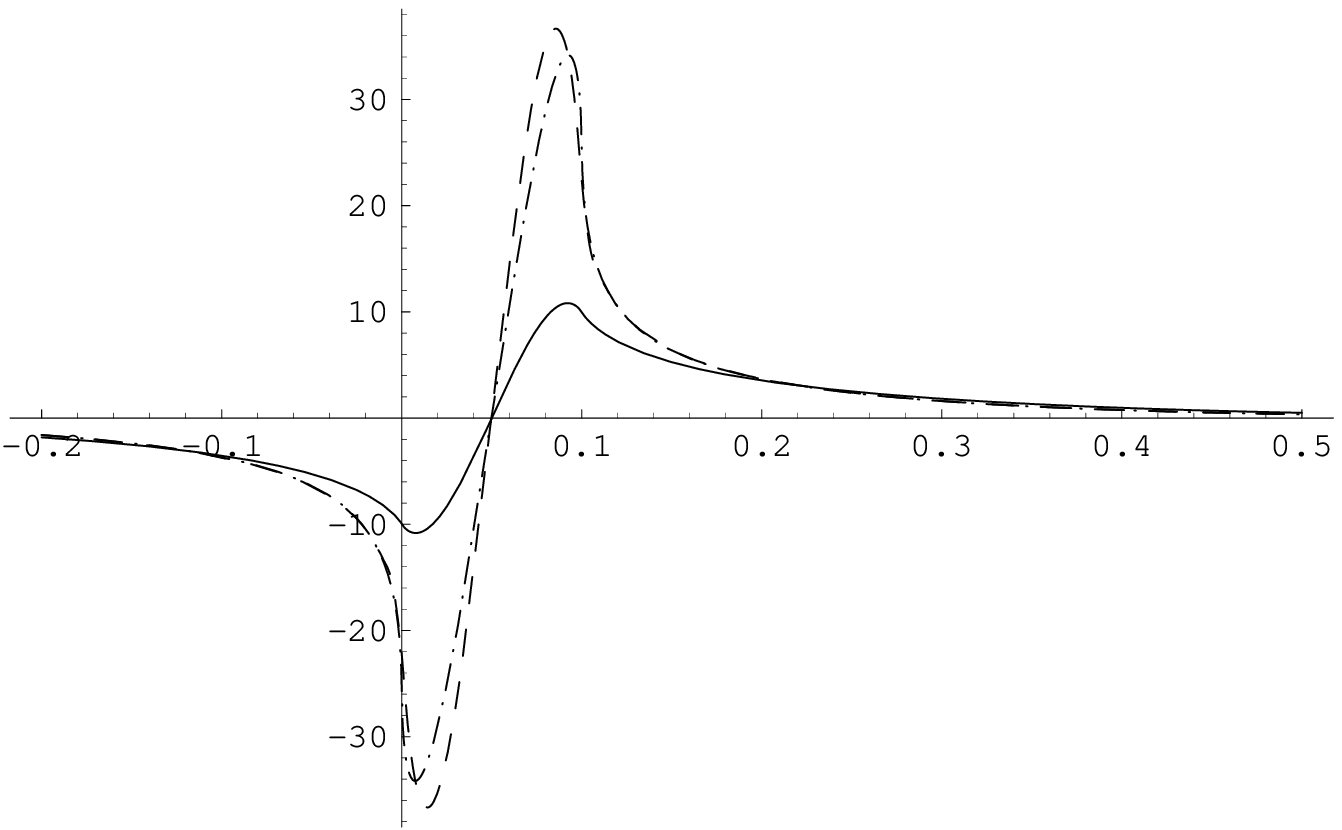}  }
{\caption{\label{Evolved-model-s}
Evolution of the QCD gluon (left) and singlet quark (right) 
nonforward distributions $F^{G,q}_\zeta(x)$
obtained our model with $h_{as}$ profiles.
Solid curves correspond to the initial distributions at $Q^2=1.5\,{\rm GeV}^2$.
Dashed curves represent nonforward distributions evolved to $Q^2=20GeV^2$,
dash-dotted ones are obtained from the model with the same  $h_{as}$ profiles
and with forward distributions evolved to $Q^2=20\,{\rm GeV}^2$.
   }}
\end{figure}

In the singlet case, it is more convenient 
to use ``tilded'' distributions defined 
on the $[-1+\zeta \leq X \leq 1]$ segment. 
These functions, shown in Fig.\ref{Evolved-model-s},  
are symmetric or antisymmetric 
wrt the middle point $X=\zeta/2$.

\section{Conclusions} 

In this paper, we discussed the structure of 
double and skewed distributions and their relation
to the usual (forward)
parton distributions. We emphasized that 
there are meson-exchange type 
terms in DDs and SPDs  which are invisible 
in the forward limit. Theoretically, they  can be modeled 
within approaches involving the mesons as 
elementary fields, e.g., this  has been  done in ref.\cite{poweiss}
within the chiral soliton
model framework.  There are also terms in SPDs
and DDs which are related to usual distributions
through reduction relations and which are,
in this sense, visible from the forward limit. 
It is just these ``forward  visible'' parts
of DDs and SPDs which are the subject of our studies in the present paper.
We proposed  factorized models for
double distributions in which one 
factor coincides with the usual (forward)
parton distribution  and another specifies the profile characterizing the
spread  of the longitudinal momentum transfer. 
By construction, these models  satisfy reduction 
formulas.  
Then we used  the  factorized model DDs   to construct 
skewed parton distributions and studied the skewedness
dependence of SPDs. We established that,   
for small skewedness, the relevant parts of 
SPDs $H(\tilde x , \xi)$ can be obtained 
by averaging  the usual parton
densities  $f(\tilde x-\xi \alpha ) $  with the weight
$\rho (\alpha)$ coinciding with the profile function
of the double distribution $\tilde f(x,\alpha)$ 
at small $x$. This result allows one to get estimates
for the ratio of SPDs taken at the border point
$\tilde x =  \xi$ (or $X=\zeta$) and 
usual parton densities taken at $x=x_{Bj}$. 
This ratio  is larger 
than 1 in all cases which we have considered,
i.e., SPDs in this sense are enhanced compared to the
 forward densities.
We found that,  for small $x_{Bj}$, the enhancement factor
is directly related to the parameter $a$ characterizing 
the effective power  behavior $x^{-a}$ of the usual 
parton densities.  We gave an explicit 
expression for the enhancement factor 
which involves $a$ and the parameter $b$ 
characterizing the effective power profile
$(1- \alpha^2)^b$ of the relevant DD.
Furthermore, we demonstrated  that if the 
$x^n$ moments $\tilde f_n (\alpha)$ of 
the $\alpha$-DDs have the asymptotic $(1-\alpha^2)^{n+1}$
profile, then the parameters $a$ and $b$ are correlated:
$b=a$, i.e., the $\alpha$-profile of $\tilde f (x,\alpha)$ 
for small $x$ is completely determined in this case 
by  the small-$x$ behavior of the usual parton distribution.
We also demonstrated that in the case of small skewedness $\xi$,
the deviation of $\tilde f_n (\alpha)$ from its asymptotic form
leads to very small $O(\xi^2)$ corrections only.
This means that the ``forward visible'' parts of SPDs
can be  rather accurately deduced from the 
purely forward usual parton densities.
The absence of higher harmonics in $\tilde f_n (\alpha)$ 
is equivalent to the absence of $\xi$-dependence 
in the Gegenbauer moments of SPDs, which is the 
starting point of the model for SPDs constructed in the recent paper
by Shuvaev et al. \cite{sgmr}.   For this reason,  the
results  based on the
asymptotic profile model (APM) developed in the present
paper coincide with those of ref. \cite{sgmr},
though its authors refrained from using DDs in their approach.
Finally, we performed a numerical investigation 
of the evolution patterns of SPDs and gave interpretation
of the results of these studies within the DD framework.
In particular, if  the initial profile
of DDs is too wide, the pQCD evolution makes it  more 
narrow and the profile widens if initially it was too narrow.
Our numerical results also support the expectation 
that if one takes SPD  derived from  the forward  distribution
by  APM prescription 
and evolves it using nonforward evolution kernels,
the result coincides with SPD which is APM-derived from evolved 
forward distribution. 
Summarizing, our results based on numerical evolution
of SPDs are in full accord with theoretical expectations
based on DD formalism.
The next step in this investigation  should be 
a direct numerical evolution of double distributions.

\acknowledgements

We thank  I. Balitsky, A. Belitsky, V. Braun, C. Coriano,
L. Frankfurt, 
 A. Freund, K. Golec-Biernat, L. Gordon, 
I. Grigentch, V. Guzey,  X. Ji, 
G. Korchemsky,  A. Martin,
L. Mankiewicz, D. M\"uller,
G. Piller,  M. Polyakov, M. Ryskin, A. Sch\"afer,
 A. Shuvaev,  M.  Strikman and 
C. Weiss for  stimulating  discussions
and  correspondence.  
This work was supported
by the US Department of Energy 
under contract DE-AC05-84ER40150.

\begin{appendix} 

\section{Asymptotic profile model
and $\xi$-independent Gegenbauer moments}

The lowest $k=0$ term of the 
$\xi^{2k}$ expansion for the Gegenbauer
moments of the SPDs involves the DDs  $\tilde f (x, \alpha)$
integrated over   $\alpha$ with $C_0^{n+3/2}(\alpha) =1$,
 i.e. the ordinary (forward)
parton distributions:
\begin{equation}
{\cal C}_{n}^Q(\xi=0, \mu) = 2^{n} \, 
 \frac{ \Gamma (n+3/2)}{\Gamma (3/2)\, n!}
  \int_{-1}^1 \tilde f^Q (x;\mu) \, x^n dx 
\  . \label{qumom}
\end{equation}
\begin{equation}
{\cal C}_{n}^G(\xi=0, \mu) = 2^n \, 
 \frac{ \Gamma (n+3/2)}{3 \, \Gamma (3/2)\, (n-1)!}
  \int_{-1}^1 \tilde f^Q (x;\mu) \, x^n dx 
\  . \label{glumom}
\end{equation}

The higher terms of the 
 $\xi^{2k}$ expansion  are small for small $\xi$.
 The approximation in which the 
  $k \neq 0$  terms  of the   $\xi^{2k}$ expansion
 for the Gegenbauer moments are neglected 
 (i.e., Gegenbauer moments ${\cal C}_{n}(\xi, \mu)$
 are treated as 
 $\xi$-independent) can be translated 
 into a model for the double distribution  $\tilde f (x, \alpha)$.
 The advantage of this model is that its
 basic assumption  is stable 
 with respect to  evolution: 
 if the Gegenbauer moments ${\cal C}_{n}(\xi, \mu_0)$
 are $\xi$-independent at some low normalization 
 point $\mu_0$, this property is preserved 
 for any higher $\mu$.  The only impact of the pQCD evolution
 is the change of the relevant forward parton distribution
 from $f(x,\mu_0)$ to $f(x,\mu)$.

To derive the explicit  expression 
for  $\tilde f (x, \alpha)$ in this model, we will use
   the expansion of the light-cone operator 
$\bar \psi(-z/2) \lambda^a \hat z \psi (z/2)$ 
over the multiplicatively renormalizable  
operators ${\cal O}_n$ (see \cite{bbal} ) 
\begin{equation}
\bar \psi(-z/2) \lambda^a \hat z \psi (z/2) =
\sum \limits_{n=0}^{\infty} (-1)^{n} 
 \frac{(2n+3)}{4^{n+1}(n+1)!}
 \int_{-1}^1 
(1-\alpha^2)^{n+1} {\cal O}_n (\alpha z/2)
\, d\alpha \ .
\label{108} \end{equation}
Inserting it into the nonforward matrix element, we obtain 
\begin{equation} 
\langle P -r/2 \,  |\,  \bar \psi(-z/2) 
\lambda^a \hat z \psi (z/2) \, 
| \, P+r/2 \,  \rangle = \int_{-1}^1 \, e^{-i\alpha (rz)/2}
\sum \limits_{n=0}^{\infty} (-1)^{n} 
 \frac{(2n+3)}{2^{n+2}(n+1)!}
  (1-\alpha^2)^{n+1}  
  (-i Pz)^n  {\cal C}_{n}(\xi, \mu) 
\, d\alpha  \  .
\label{1080} 
\end{equation}

If the Gegenbauer moments ${\cal C}_{n}(\xi, \mu)$
are approximated by their $\xi =0 $ values,
Eq.(\ref{1080}) can be  transformed 
into the representation of the nonforward matrix element
in terms of double distributions.
Namely, the $n$th moment 
\begin{equation}
\tilde f_n(\alpha \,  ;     \,  \mu) = \int_{-1}^{1} x^n  
\tilde f(x,\alpha\,  ;     \,  \mu) \, dx \, 
\label{eq:fnmom}
 \end{equation}
of the $\alpha$-DD $\tilde f(x,\alpha\,  ;     \,  \mu)$
is then given by 
\begin{equation} 
\tilde f_n (\alpha \,  ;     \,  \mu)
=   
 \frac{\Gamma(n+5/2)}{\Gamma (1/2) \, (n+1)!}
  (1-\alpha^2)^{n+1}  
  \int_{-1}^{1} f (z;\mu) z^n dz \  .
\label{1081} 
\end{equation}
 The factor 
relating $\tilde f_n (\alpha \,  ;     \,  \mu)$ 
and $\tilde f_n (     \,  \mu)$ is just 
the  normalized profile function $\rho_{n+1}(\alpha)$ (see Section ?). 
 In the  nonsinglet case only 
 moments with even $n$ are  nonzero,
 while in the singlet case 
only those with odd $n$ do not vanish. 
 Below, we will  construct the $x>0$   parts of DDs, 
the trivial (anti)symmetrization can be performed at the end.
 
Incorporating the inverse Mellin transformation,
one can obtain the kernel $K(x,\alpha ; z)$ which relates 
DDs $ f(x,\alpha)$ in this model with the 
usual forward distributions $ f(z)$.
However, inverting  the representation (\ref{1081})  ``as is'',
one would get the expression 
\begin{equation} 
 f (x, \alpha) \stackrel{?}{=} -
\frac1{2\pi} \int \limits_{x/ (1-\alpha^2)}^1  \,    
[z (1-\alpha^2)/x-1 ]^{-3/2} \,
  f(z) \, \frac{dz}{z}
 \ ,   
\end{equation} 
 whose rhs  has  a  suspicious overall sign.
 Furthermore,  
 the  integral over $z$ diverges 
at the end-point $z= x/(1-\alpha^2)$. These 
inconsistencies      
indicate that the implied interchange of the inverse Mellin 
transformation and the $z$-integral  is not justified. 
To get a   less singular kernel, one 
can try to  add an $O(1/n)$ factor in the expression 
for the moments  $f_n (\alpha \,  ;     \,  \mu)$,
e.g., convert $1/(n+1)!$ into $1/(n+2)!$
(after such a change, the inverse Mellin transform
is still easily doable analytically).
To this end, we use the relation
\begin{equation} 
\int_{0}^{1} f (z;\mu) z^n dz  = - \frac1{n+2} 
\int_{0}^{1} z^{n+2} \,\frac{d}{dz} 
\left ( \frac{f (z;\mu)}{z} \right )   dz \ 
\label{trick}
\end{equation}
which holds if $(a)$ the  function $f (z;\mu)$ 
vanishes at $z =  1$ (this is always true)
 and $(b)$  $z^{n+1} f(z) $ 
vanishes at $z=0$. The latter requirement
is evidently  satisfied in the nonsinglet case
for all $n \geq 0$.     
The singlet distributions are more singular,
but we  need  only their $n \geq 1$ moments, i.e., 
the restriction  $(b)$ is  satisfied again.  
Using Eq.(\ref{trick})   
produces  the  kernel which 
connects $ f(x,\alpha)$ 
with $[ f(z)/z]^{\prime}$:
\begin{eqnarray}
 f (x, \alpha) = - \frac{x}{\pi (1- \alpha^2)}
\int \limits_{x/ (1-\alpha^2)}^1   
\left  [z(1-\alpha^2)/x -1 \right ]^{-1/2} 
   \left [ f(z)/z \right ]^{\prime} \,  
 dz 
 \ .  
 \end{eqnarray}
This result coincides with  Eq.(\ref{inabel}). 
The spectral condition $x/z \leq 1-\alpha^2$ 
 relating the ``original'' fraction $z$ and the 
 ``produced''  fraction $x$ is analogous to the momentum ordering
 $x\leq z$ in the DGLAP equation: the produced 
 fraction cannot be larger than the original one.
  In the present case, if the parton also takes some 
  nonzero fraction 
 $\alpha$ of the momentum transfer, the allowed 
 values of $x$ cannot exceed $z (1- \alpha ^2)$.

For gluons, we combine  the expansion for  
the bilocal operator  
 \begin{equation}
   z^{\mu} z^{\nu} G_{\mu \alpha}(-z/2)
 G_{\alpha \nu}(z/2)= 
\sum \limits_{n=0}^{\infty} (-1)^{n} 
 \frac{3 (2n+5)}{2^{2n+3}(n+2)!}
 \int_{-1}^1 
(1-\alpha^2)^{n+2} {\cal O}_{n+1}^G (\alpha z/2)
\, d\alpha 
\label{1080g} \end{equation}
and the expression (\ref{glumom}) for the
Gegenbauer moments at $\xi =0$. 
The resulting relation for the moments
$\tilde f_n^G (\alpha ; \mu)$ has the  form
identical to Eq.(\ref{1081})  derived in the quark case.

\section{Evolution Equations}

In Ref. \cite{npd}, the kernels $W_\zeta^{ab}(x,z)$ 
were obtained $via$ 
\begin{eqnarray}
W_\zeta^{ab} (x,z) &=&
  \int_0^1 \int_0^1  B^{ab}(u,v)  
   \delta\left (x-\bar uz+v(z-\zeta)\right ) 
 \theta(u+v\leq 1) \ du\ dv
\label{Wab}          
\end{eqnarray}
 from the evolution kernels for 
the  light--ray operators \cite{bbal,brschwg}:
\begin{eqnarray}
B^{qq} &=& {\alpha_s\over \pi} C_F
            \left ( 1+\delta(u) [\bar v/v]_++\delta(v) [\bar u/u]_+ -
            {1\over 2} \delta(u)\delta(v) \right ) 
\label{Bqq}  \\
B^{gq} &=& {\alpha_s\over \pi} C_F
            \left ( 2+\delta(u) \delta(v)  \right )
\label{Bgq}  \\
B^{qg} &=& {\alpha_s\over \pi} N_f
            ( 1+4uv-u-v )
\label{Bqg}  \\
B^{gg} &=& {\alpha_s\over \pi} N_c
            \left ( 4(1+3uv-u-v)
                  + {\beta_0\over 2N_c} \delta(u) \delta(v)
                  +  \delta(u) {\bar v}^2 \left [{1\over v}\right ]_+
                  +  \delta(v) {\bar u}^2 \left [{1\over u}\right ]_+
            \right ) 
\label{Bgg}           
\end{eqnarray}

In the nonsinglet case, only the
$qq$ kernel is needed.
Using Eqs. (\ref{Bqq}),(\ref{Wab}), one can easily derive
the rules allowing to  transform  each 
of  the four terms contained  
inside the bracket in Eq. (\ref{Bqq})   
\begin{eqnarray}
     1 & \to &
       \theta(x < \zeta)
       \int\limits_0^x 
                {\zeta - x \over \zeta - z} \,  {\cal F}_\zeta^q(z)
		{dz \over \zeta}
     + \int\limits_x^1  \left \{
        \theta(x < \zeta) 
                         \, {x \over \zeta} {dz \over z}
     +  \theta(x > \zeta) \, 
           {z - x \over z - \zeta} \right \} \, {\cal F}_\zeta^q(z) 
	   \equiv d_1, \label{d1} \\
    \delta(u) \left [\frac{\bar v}{v}\right ]_+ &\to & 
        \int\limits_0^{\min\{x/\zeta,\bar x/\bar \zeta\}} 
                \left [ {\cal F}_\zeta^q \left ({x - v \zeta \over \bar v} \right )
                    - {\cal F}_\zeta^q(x) \right ] \, {dv \over v}+
        \left [1 + \log( \min\{x/\zeta,\bar x/\bar \zeta\} )\right ] 
           {\cal F}_\zeta^q(x)  \nonumber \\
	&=& \theta(x < \zeta) \left \{
        \int\limits_0^{x/\zeta} 
                \left [ {\cal F}_\zeta^q \left ({x - v \zeta \over \bar v} \right )
                    - {\cal F}_\zeta^q(x) \right ]\, {dv \over v} +
        \left [1 + \log(x/\zeta)\right ]  
           {\cal F}_\zeta^q(x) \right \} \nonumber\\
	&+& \theta(x > \zeta) \left \{
        \int\limits_0^{\bar x/\bar \zeta} 
                \left [ {\cal F}_\zeta^q \left ({x - v \zeta \over \bar v} \right )
                    - {\cal F}_\zeta^q(x) \right ]\, {dv \over v} +
        \left [1 + \log(\bar x/\bar \zeta)\right ]
           {\cal F}_\zeta^q(x)  \right \} \equiv d_2  , \label{d2} \\	
   \delta(v) \left [\frac{\bar u}{u}\right ]_+  &\to & 
      \int\limits_0^{\bar x} 
      \left [ {\cal F}_\zeta^q(x/\bar u) -  {\cal F}_\zeta^q(x)\right ]\, {du \over u} + 
        \left (1 + \log(\bar x)\right ) {\cal F}_\zeta^q(x) \equiv d_3 , \label{d3}\\
           -{1\over 2} \delta(u)\delta(v)  & \to & -{1\over 2}
	     {\cal F}_\zeta^q(x) \equiv d_4 \, .\label{d4}
	     \label{NFkernqq}
\end{eqnarray}
In terms of these contributions, the total $qq$ part of 
(\ref{EvEqn}) is given by
\begin{equation} 
 \hat W_\zeta^{qq}
 \otimes  \hat{\cal F}_\zeta^q(\mu) 
      = {\alpha_s \over \pi} C_F (d_1 + d_2 + d_3 + d_4).
\end{equation}

Expressions for $d_1$ and $d_4$ were obtained directly by performing
integrations over $u$, $v$ of the terms in Eq. (\ref{Bqq}) with $1$ and
$(-\frac12 \delta(u)\delta(v))$. To obtain expressions for $d_2$ and $d_3$,
we changed the order of integrations.
For example, to obtain $d_3$, we first take the integral
over $v$ and  then over $z$:
\begin{eqnarray}
&&\int_0^1 \left ( \int_0^1\int_0^1 \delta(x-\bar uz+v(z-\zeta))
  \delta(v)\left [{\bar u\over u}\right ]_+ {\cal F}_\zeta(z)\ du\ dv 
  \right ) dz = \nonumber\\
&&\int_0^1 \left ( \int_0^1 \delta(x-\bar uz)
                 \left [{\bar u\over u}\right ]_+ {\cal F}_\zeta(z)\ du \right ) dz
   = \int_0^1 {du\over \bar u} \theta(x/\bar u <1) 
     \left [{\bar u\over u}\right ]_+ {\cal F}_\zeta(x/\bar u) \nonumber\\
&& = \int_0^{\bar x} {\cal F}_\zeta(x/\bar u)\, {du\over u} -
{\cal F}_\zeta(x) \int_0^1  {\bar u\over u}\, du \nonumber\\
&& =       \int\limits_0^{\bar x} {du \over u}
      \left [ {\cal F}_\zeta^q(x/\bar u) -  {\cal F}_\zeta^q(x)\right ] + 
        \left (1 + \log(\bar x)\right ) {\cal F}_\zeta^q(x). 
\end{eqnarray}
The resulting   form for integrals in $d_2$ and $d_3$ 
is particularly convenient  for  numerical calculations, because
the integrand in Eqs. (\ref{d2}), (\ref{d3}) is explicitly regular as $u,v \to 0$.

In a similar way, we represent the    $gg$ -kernel part as 
\begin{equation} 
 \hat W_\zeta^{gg}
 \otimes  \hat{\cal F}_\zeta^g(\mu^2) 
      = {\alpha_s \over \pi} N_c (d_1^g + d_2^g + d_3^g + d_4^g).\
\end{equation}
where 
\begin{eqnarray}
d^g_1 &=&
       2\,  \theta(x < \zeta) \left \{
       \int\limits_0^x  {\zeta-x \over \zeta^2 (\zeta-z)^2} \
               \left ( \zeta(2+x-z) -x^2 -z(2-x) \right )  
               {\cal F}_\zeta^g(z)\, dz
     + \int\limits_x^1 \, 
                         {x^2 \over \zeta^2}
                    \left ( 3 - {2x\over\zeta} + {\zeta-x\over z} \right )
                         {\cal F}_\zeta^g(z){dz \over z}
                           \right \} \nonumber \\
    &+&  2 \, \theta(x > \zeta) \int\limits_x^1 
                         {z-x \over z-\zeta}
                    \left ( 1 + {x(x-\zeta)\over z(z-\zeta)} \right )
                         {\cal F}_\zeta^g(z)\, {dz \over z}
                     \\
d^g_2 &=&
        \theta(x < \zeta) \left \{
        \int\limits_0^{x/\zeta} 
                \left [\bar v{\cal F}_\zeta^g \left ({x - v \zeta \over \bar v} \right )
                    - {\cal F}_\zeta^g(x) \right ]\, {dv \over v} +
        \log(x/\zeta) \, {\cal F}_\zeta^g(x) \right \}  \nonumber\\
	&+& \theta(x > \zeta) \left \{
        \int\limits_0^{\bar x/\bar \zeta} 
                \left [\bar z{\cal F}_\zeta^g \left ({x - v \zeta \over \bar v} \right )
                    - {\cal F}_\zeta^g(x) \right ]\, {dv \over v} +
        \log(\bar x/\bar \zeta)
           {\cal F}_\zeta^g(x)  \right \} , \label{dg2} \\	
d^g_3 &=& 
      \int\limits_0^{\bar x} 
      \left (\bar u{\cal F}_\zeta^g(x/\bar u) -  {\cal F}_\zeta^g(x)\right )
      \, {du \over u} + 
      \log(\bar x) {\cal F}_\zeta^g(x), \label{dg3}\\
d^g_4 &=&
        {\beta_0 \over 2N_c}  {\cal F}^g_\zeta (x) \label{dg4} \label{NFkerngg}
\end{eqnarray}

The $qg$ and $gq$ parts  of the evolution equations
(\ref{formalEvEqn}) for untilded NFPDs 
in the region where $z < \zeta$ cannot be 
unambiguously  reconstructed from the light-ray kernels. 
We use the form suggested in \cite{belmul}, which leads to
\begin{eqnarray}
s^{gq} &=&
         \theta(x > \zeta)\, {\zeta-x\over\zeta}\int\limits_x^1 
                         {z-x\over z-\zeta}
                    \left ( 1 + {x(x-\zeta)\over z(z-\zeta)} \right )
                         {\cal F}_\zeta^q(z)\, {dz\over z}  \\
       &+& 
       \theta(x < \zeta) \left \{
        {(\zeta-x)^2\over\zeta}
        \int\limits_0^x {1\over \zeta-z}
          \left ( 1- {2z\over\zeta} + 4\, {x (\zeta-z)\over\zeta^2} \right )
               {\cal F}_\zeta^q(z)\, dz   \right . \\
        &&+ 
         \left . {x^2\over\zeta^2}
         \int\limits_x^1 
         \left ( -1 + {2z\over\zeta} + 4{z (\zeta-x)\over\zeta^2} \right )
                         {\cal F}_\zeta^q(z)\, dz 
                           \right \},
\label{NFkerngq} \\
s^{qg} &=&
         \theta(x > \zeta) \int\limits_x^1 
                         {(z-x)^2+x(x-\zeta)\over z^2 (z-\zeta)^2}
                    \left ( 1 + {x(x-\zeta)\over z(z-\zeta)} \right )
                         {\cal F}_\zeta^q(z) \, dz \\
        &+& 
       \theta(x < \zeta) \left \{
        {-(\zeta-x)\over\zeta}
        \int\limits_0^x 
         {1\over (\zeta-z)^2} 
          \left ( 1- {2x\over\zeta} - 4{x (\zeta-z)\over\zeta^2} \right )
               {\cal F}_\zeta^g(z) \, dz \right . \\
        &&+ \left . {x\over\zeta}
         \int\limits_x^1 {1\over z^2}
         \left ( -1 + {2x\over\zeta} - 4{z (\zeta-x)\over\zeta^2} \right )
                         {\cal F}_\zeta^g(z) \, dz 
                           \right \}.
\label{NFkernqg} 
\end{eqnarray}
In the matrix notation
\begin{eqnarray} 
&&  \hat W_\zeta^{gq}
 \otimes  \hat{\cal F}_\zeta^q(\mu^2) 
      = {\alpha_s \over \pi}  C_f s^{gq}, 
\label{NFkerngq-formal}\\
&&  \hat W_\zeta^{qg}
 \otimes \hat{\cal F}_\zeta^g(\mu^2) 
      = {\alpha_s \over \pi} N_f s^{qg}.
\label{NFkernqg-formal}
\end{eqnarray}

\end{appendix}

\end{document}